\documentclass[notitlepage,nofootinbib,preprintnumbers,amssymb,groupedaddress]{revtex4-1}
\usepackage{amsfonts,amssymb,amsmath,graphicx,color,bm,natbib}
\definecolor{ultramarine}{rgb}{0.07, 0.04, 0.56}
\definecolor{cadmiumgreen}{rgb}{0.0, 0.42, 0.24}
\definecolor{indigo(dye)}{rgb}{0.0, 0.25, 0.42}
\usepackage[linktocpage=true]{hyperref}
\hypersetup{
colorlinks=true,
citecolor=ultramarine, 
linkcolor=red, 
urlcolor=indigo(dye),
}

\newcommand{\be}{\begin{eqnarray}}  
\newcommand{\ee}{\end{eqnarray}}
\newcommand{\bem}{\begin{bmatrix}}
\newcommand{\eem}{\end{bmatrix}}

\allowdisplaybreaks

\begin{document}

\title{Spontaneous scalarization of a conducting sphere in Maxwell--scalar models}

\author{Carlos A. R. Herdeiro$^{1,}$\footnote{herdeiro@ua.pt}, Taishi Ikeda$^{2,3}$\footnote{taishi.ikeda@tecnico.ulisboa.pt}, Masato Minamitsuji$^{2,}$\footnote{masato.minamitsuji@ist.utl.pt}, Tomohiro Nakamura$^{4,}$\footnote{nakamura.tomohiro@g.mbox.nagoya-u.ac.jp} and Eugen Radu$^{1,}$\footnote{eugen.radu@ua.pt} \ \ \ }
\affiliation{${^1}$Departamento de Matem\'atica da Universidade de Aveiro and \\ Centre for Research and Development  in Mathematics and Applications (CIDMA), Campus de Santiago, 3810-183 Aveiro, Portugal }
\affiliation{${^2}$ Centro de Astrof\'{\i}sica e Gravita\c c\~ao  - CENTRA, Departamento de F\'{\i}sica, Instituto Superior T\'ecnico - IST, Universidade de Lisboa - UL, Av. Rovisco Pais 1, 1049-001 Lisboa, Portugal}
\affiliation{${^3}$Dipartimento di Fisica, ``Sapienza'' Universit\'{a} di Roma, Piazzale Aldo Moro 5, 00185, Roma, Italy}
\affiliation{${^4}$Department  of  Physics,  Nagoya  University,  Nagoya  464-8602,  Japan }


\date{\today}

\begin{abstract}
We study the spontaneous scalarization of a standard conducting charged sphere embedded in Maxwell-scalar models in flat spacetime, wherein the scalar field $\phi$ is nonminimally coupled to the Maxwell electrodynamics. This setup serves as a toy model for the spontaneous scalarization of charged (vacuum) black holes in Einstein-Maxwell-scalar (generalized scalar-tensor) models. In the Maxwell-scalar case, unlike the black hole cases, closed-form solutions exist for the scalarized configurations. We compute these configurations for three illustrations of nonminimal couplings: one that \textit{exactly} linearizes the scalar field equation, and the remaining two that produce nonlinear continuations of the first one. We show that the former model leads to a runaway behaviour in  regions  of the parameter space and neither the Coulomb nor the scalarized solutions are stable in the model; but the latter models can heal this behaviour producing stable scalarized solutions that are dynamically preferred over the Coulomb one. This parallels reports on black hole scalarization in the extended-scalar-Gauss-Bonnet models. Moreover, we analyse the impact of the choice of the boundary conditions on the scalarization phenomenon. Dirichlet and Neumann boundary conditions accommodate both (linearly) stable and unstable parameter space regions, for the scalar-free conducting sphere; but radiative boundary conditions always yield an unstable scalar-free solution and preference for scalarization. Finally, we perform numerical evolution of the full Maxwell-scalar system, following dynamically the scalarization process. They confirm the linear stability analysis and reveal that the scalarization phenomenon can occur in qualitatively distinct ways.
\end{abstract}




\maketitle

\tableofcontents

\section{Introduction}
\label{sec1}
Compact objects such as neutron stars and black holes
can undergo a spontaneous scalarization phenomenon in asymptotically flat spacetimes, in scalar-tensor
theories of gravitation, 
see e.g.~\cite{Damour:1993hw,Damour:1996ke,Harada:1997mr,Harada:1998ge,Novak:1998rk,Silva:2014fca,Motohashi:2018mql}
for scalarization of relativistic stars 
and 
e.g., 
\cite{Silva:2017uqg,Doneva:2017bvd,Antoniou:2017acq,Doneva:2017duq,Blazquez-Salcedo:2018jnn,Silva:2018qhn,Bakopoulos:2018nui,Antoniou:2017hxj,Doneva:2018rou,Minamitsuji:2018xde,Brihaye:2018grv,Brihaye:2018bgc,
Doneva:2019vuh,Cunha:2019dwb,Macedo:2019sem,Herdeiro:2019yjy,Collodel:2019kkx,Brihaye:2019gla,Hod:2019pmb,Peng:2019qrl,Hod:2019vut,Minamitsuji:2019iwp,Blazquez-Salcedo:2020rhf,Blazquez-Salcedo:2020caw,Doneva:2020qww,Astefanesei:2020qxk,Peng:2020znl,Dima:2020yac,Hod:2020jjy,Doneva:2020nbb,Doneva:2020kfv,Herdeiro:2020wei,Berti:2020kgk}
for black hole scalarization.
This is a strong gravity phase transition, wherein the nonminimal coupling between the scalar field 
and a curvature invariant (or matter density) plays a key role
that triggers a tachyonic growth of the scalar field in high curvature (or high density) backgrounds.

\smallskip

Spontaneous scalarization, however, is not exclusive of either scalar-tensor models or even of strong gravity\footnote{Also, a similar phenomenon exists for vector and tensor (rather than scalar) fields, see and e.g.~\cite{Annulli:2019fzq,Ramazanoglu:2019gbz,Ramazanoglu:2019jrr,Kase:2020yhw,Minamitsuji:2020pak}.}; it occurs in a broader class of models. For instance, if the scalar field nonminimally couples to the Maxwell (rather than a curvature) invariant $F^{\mu\nu}F_{\mu\nu}$, with $F_{\mu\nu}$ being the electromagnetic field strength, charged black holes can scalarize in Einstein-Maxwell-scalar models~\cite{Herdeiro:2018wub,Myung:2018jvi,Myung:2018vug,
Fernandes:2019rez,Fernandes:2019kmh,Myung:2019oua,Zou:2019bpt,Brihaye:2019kvj,Ikeda:2019okp,
Fernandes:2020gay,Hod:2020ljo,Hod:2020ius}, 
wherein both the scalar and electromagnetic fields \textit{minimally} couple to Einstein's gravity (see also \cite{Stefanov:2007qw,Stefanov:2007eq} for scalar field couplings in nonlinear electrodynamics). 
Additionally, even in the absence of gravity, 
a nonminimal coupling of the scalar field to Maxwell's electromagnetism, 
$f(\phi)F^{\mu\nu}F_{\mu\nu}$,
where $f(\phi)$ is a regular function of $\phi$,
may allow scalarization of a charged object, such as a conducting sphere~\cite{Herdeiro:2018wub}
(see also~\cite{Hod:2020fht}). 
This occurs in Maxwell-scalar (Ms) models on Minkowski spacetime, 
which therefore provide one of the simplest arenas to study the spontaneous scalarization phenomenon. 
Studying this phenomenon in detail will be the subject of this paper. 

\smallskip

Ms models are rather simple for the static and spherically symmetric configurations,
which 
possess an integrable structure~\cite{Herdeiro:2019iwl}.
Then, despite being non-trivially coupled to the electromagnetic dynamics via a nonminimal coupling function $f(\phi)$, the scalar field equation can be fully decoupled from Maxwell's equations via considering the first integral that arises from the latter,  the electric charge. Consequently, specifying $f(\phi)$, one can fully grasp the linear/nonlinear structure of the scalar field equation. 
This is unlike what happens in scalar-tensor  or Einstein-Maxwell-scalar models, 
in which the impact of the geometry on the scalar field equation 
and the precise structure of the corresponding nonlinearities 
are not so explicit in general, since the geometry is sourced by the scalar field.

\smallskip

The first investigation of spontaneous scalarization in an Ms model was briefly mentioned in~\cite{Herdeiro:2018wub}, wherein a closed-form solution describing a scalarized conducting sphere was reported. 
This was found for the specific nonlinear form of $f(\phi)$ given by Eq.~\eqref{model1} below. 
As it turns out, the particular simplicity induced by this coupling can be traced to the fact that it \textit{exactly} linearizes the scalar field equation. Using appropriate variables, the latter is, in fact, simply the standard harmonic equation. In~\cite{Herdeiro:2018wub}, it was observed that the corresponding scalarized conductor was energetically preferred over the Coulomb solution with the same charge in a subset of the parameter space,
but no dynamical or stability analysis was performed in Ref.~\cite{Herdeiro:2018wub}.
Such an analysis will be performed 
in this paper, leading to the conclusion  that this model is afflicted by a runaway mode: 
there are regions of the parameter space wherein neither the scalar-free (Coulomb) 
nor the scalarized conducting sphere are stable. 

\smallskip

A physically more reasonable behaviour, with a stable (classical) vacuum, 
can be obtained by augmenting the previous model with sufficient nonlinearities. 
This will be illustrated by considering two examples of nonlinear couplings $f(\phi)$ 
given by Eqs.~\eqref{model2} and~\eqref{model3} below, 
that can be regarded as nonlinear continuations of the first choice that led to the  linear model. 
In fact, the scalar field equation now  becomes  that of a nonlinear  harmonic oscillator. 
The nonlinearities that these couplings induce in the latter equation will be shown 
to be able to quench the runaway mode, by introducing an effective potential with new extrema, 
yielding stable  scalarized configurations. 

\smallskip

This dichotomy 
between the linear and nonlinear models and (non)existence of a stable scalarized vacuum,
which was just explained above, 
finds a close parallelism with the situation found in black hole spontaneous scalarization 
in extended-scalar-tensor-Gauss-Bonnet models. 
The simplest nonminimal coupling allowing for black hole scalarization~\cite{Silva:2017uqg}, 
which produces a linear scalar field equation\footnote{This linearity is only apparent, 
since the scalar field backreacts on the geometry which then impacts on the scalar field equation.},  
yields static scalarized black holes that are not entropically preferred over the Schwarzschild solution with the same mass and, moreover, which were later found to be radially unstable~\cite{Blazquez-Salcedo:2018jnn}. 
But couplings which produces a manifestly nonlinear scalar field equation 
could give rise to perturbatively stable scalarized solutions~\cite{Doneva:2017bvd,Blazquez-Salcedo:2018jnn,Silva:2018qhn,Minamitsuji:2018xde,Blazquez-Salcedo:2020rhf,Blazquez-Salcedo:2020caw}.
We suggest that the aforementioned Ms models, described herein, are a simple toy model 
to study this behavior and the dynamical formation process of scalarized solutions.

\smallskip

In the Ms models, unlike the black hole case, however,
there is a larger freedom in choosing the boundary conditions for the scalar field
on the surface of the conducting sphere.
We will show that the choice of the boundary conditions can impact significantly on the instability 
of the conducting  sphere.
These  findings will result from a linear perturbation  theory analysis, 
which is also performed for the scalarized solutions, but will also be confirmed by fully nonlinear numerical evolution. 
Moreover, 
the latter will help us understand \textit{how} the scalarized solution is reached. 
We will see that the tachyonic scalarization instability is always stronger in the immediate vicinity 
of the surface of the conducting sphere,
and then propagates towards large radial distances.
Moreover, in some cases the scalarization instability proceeds in a rather monotonic 
and straightforward manner to form the scalarized solutions, 
whereas in other cases the scalar  field oscillates considerably before approaching the end point.

\smallskip

This paper is organized as follows. Ms models are briefly described in Section~\ref{sec2}, where the Coulomb solution and its stability, when embedded in these models, is considered for different boundary conditions.  Section~ \ref{sec3} describes the  \textit{linear} model  and its scalarized solution, establishing they are always  unstable. Section~ \ref{sec4} describes the two \textit{nonlinear} models  and their scalarized solution, establishing the regimes in which they are stable. In Section~\ref{sec_num_sim}  we describe the numerical  simulations  to study the time evolution of the scalarization phenomenon. Some conclusions are presented in Section~\ref{sec6}.  Two appendices  provide some further technical details.

\section{Maxwell-scalar (M{\tiny{s}}) models}
\label{sec2}

\subsection{Action and field equations}
Let us consider the following family of Ms models,  wherein a real scalar  field $\phi$ is nonminimally coupled  to the electromagnetic field:\footnote{ The standard normalizations of the scalar and Maxwell kinetic terms correspond to multiplying this action by $1/4$.}
\be
\label{action}
S=
-\int d^4 x
\left[
2 \partial^\mu \phi \partial_ \mu \phi
+ f(\phi) F^{\mu\nu}F_{\mu\nu}
\right] \ ,
\ee
where 
$F_{\mu\nu}=\partial_\mu A_\nu- \partial_\nu A_\mu$ is the electromagnetic field strength, 
$A_\mu$ is the $U(1)$ gauge field,
and 
$f(\phi)$ is a free function of the scalar field $\phi$, specifying the nonminimal coupling. 
For $f=1$ 
the scalar field is free and it is not sourced by the electromagnetic field; 
we shall be interested in non-trivial functions $f(\phi)$
that allow a scalarization phenomenon,  triggered by charged configurations.
Throughout this work, dynamics will take place on flat Minkowski spacetime, 
whose metric is taken in the standard spherical coordinates 
\be
ds^2
=
-dt^2
+dr^2
+r^2
\left(
  d\theta^2
+\sin^2\theta d{\bar \varphi}^2
\right),
\ee
where 
$t$, $r$, and $(\theta,{\bar\varphi})$ are the time, radial and angular coordinates,
respectively.
By varying the action \eqref{action}
with respect to $\phi$ and $A_\mu$,
we obtain the Ms field equations:
\be
&&
\label{eq1}
\Box\phi
-\frac{1}{4}f_\phi(\phi) F^{\mu\nu}F_{\mu\nu}
=0 \ ,
\\
&&
\label{eq2}
D_\mu 
\left(
f(\phi)F^{\mu\nu}
\right)
=0 \ ,
\ee
where $f_\phi:=d f/d\phi$. We shall be interested in spherical dynamics and spherical static configurations. As such, for static solutions we assume a spherically symmetric ansatz for both the gauge potential and the scalar field: 
\be
\label{san}
&&
A_\mu
=(A_t,A_r,A_\theta,A_{\bar \varphi})
:=(A_0(r),0,0,0)\ ,
\\
&&
\phi=\phi_0(r) \ .
\ee
In this spherical sector, the Ms model is integrable~\cite{Herdeiro:2019iwl}. 
The nontrivial components of the vector and scalar field equations of motion, 
Eqs. \eqref{eq1} and \eqref{eq2}, 
 are given by 
\begin{eqnarray}
&&
\label{seq}
\phi_0''(r)
+\frac{2}{r} \phi_0'(r)
+\frac{1}{2}
f_\phi(\phi_0) A_0'(r)^2
=0 \ ,
\\
&&
\label{eeq}
A_0''(r)
+\frac{2}{r}A_0'(r)
+\frac{f_\phi(\phi_0)}
         {f(\phi_0)}
\phi_0'(r)
A_0'(r)
=0 \ ,
\end{eqnarray}
where `prime' denotes radial derivative.
Integrating the second equation yields a first integral:
\begin{eqnarray}
\label{electric}
A_0'(r)
=\frac{Q}{r^2 f(\phi_0)} \ ,
\end{eqnarray}
where the integration constant $Q$ is the total electric charge.
The scalar field equation \eqref{seq} then reduces to 
\be
\label{static_eq}
r^2\frac{d}{dr}\left(r^2\frac{d\phi_0}{dr}\right)=\frac{d}{d\phi_0}\left(\frac{Q^2}{2f(\phi_0)}\right) \ .
\ee
Defining a new radial coordinate $x:=1/r$ and  an  effective potential for the motion
\be
\mathcal{V}_{\rm eff}(\phi_0):=-\frac{Q^2}{2f(\phi_0)} \ ,
\label{potmot}
\ee
Eq.~\eqref{static_eq} becomes
\be
\frac{d^2\phi_0}{dx^2}=-\frac{d\mathcal{V}_{\rm eff}}{d\phi_0} \ .
\label{xeq}
\ee
This describes a mechanical model of a particle with degree of freedom $\phi_0$ moving under the influence of a 1-dimensional potential~\eqref{potmot}, with the inverse radius $x$ playing the role of `time'. This potential is, up to a constant,  the inverse of the coupling function. 
The model can be solved for any choice of the nonminimal coupling function $f(\phi)$. We remark that a second integral can be obtained from~\eqref{static_eq}~\cite{Herdeiro:2019iwl}.

\subsection{The Coulomb solution embedded in Ms models and its stability}
\label{sec2b}
The Coulomb solution in standard electromagnetism, plus a trivial scalar field,
\be
A_0'(r)=\frac{Q}{r^2 } \ ,
\qquad 
\phi_0(r)=0 \ ,
\ee
solve the Ms theory \eqref{action} as  long as the coupling function obeys
\be
\label{phizero}
f(0)=1 \ ,
\qquad 
f_\phi(0)=0 \ .
\ee
If $f(\phi)$ admits a Taylor series around $\phi=0$, then the general form of the coupling satisfying \eqref{phizero}
is given by 
\begin{eqnarray}
\label{general}
f(\phi)
=1+a \phi^2 +\sum_{n=3}^\infty f_n \phi^{n} \ ,
\end{eqnarray}
where $a>0$ and $f_n$ ($n=3,4,\cdots$) are constants.

Embedding the Coulomb solution in a Ms theory raises the question {\it whether it is the preferred vacuum in spherical symmetry with charge $Q$. Or, could there be another dynamically preferred vacuum? And, in that case, do scalar perturbations  promote the evolution of the Coulomb solution towards the preferred vacuum? }

In order to address these questions,  the physical setup herein is to consider a conducting  charged sphere with a radius $r=r_{\rm s}$. This circumvents the well-known issue in classical electromagnetism that the energy of the Coulomb solution diverges as $r\rightarrow 0$. Having such a sphere, the Coulomb solution holds on and outside the conductor. In our study, we shall always focus on the behaviour on and outside the conductor only.

In Appendix \ref{appendix1},
we discuss the linear radial perturbations in general backgrounds in Ms models.
For the coupling given by Eq. \eqref{general}, 
scalar and vector field perturbations decouple in Eq. \eqref{scalar_pert_eq};
the scalar field radial perturbation $\phi_1(r)$ obeys
\begin{eqnarray}
-\ddot{\phi}_1
+\phi_1''
+\frac{2}{r} \phi_1'
+ \frac{aQ^2}{r^4}\phi_1
=0 \ ,
\end{eqnarray}
where `dot' stands for a time derivative.
Fourier decomposing the solution,
$
\phi_1(t,r)= \int d\omega e^{-i\omega t} \phi_\omega(r)$,
each mode obeys 
\be
 \phi_\omega''(r)
+\frac{2}{r} \phi_\omega'(r)
+ \frac{aQ^2}{r^4}\phi_\omega(r)
=
-\omega^2\phi_\omega(r) \ .
\ee

An instability is present if there are modes 
with purely imaginary eigenvalues, $\omega^2<0$
(see Appendix \ref{appendix2} for the proof).  
Thus, instabilities are  absent if the mode $\omega^2=0$ is the lowest eigenmode; 
in this case, the solution $\phi_{\omega=0}$ has no nodes
between the conductor at $r=r_{\rm s}$
and spatial infinity $r\to \infty$.
For convenience, we introduce
the dimensionless quantities,
\be
\label{barred_quantities}
\bar{r}:=\frac{r}{r_{\rm s}} \ , \qquad \bar{Q}:=\frac{Q\sqrt{a}}{r_{s}},
\ee
so that the conductor surface is located at $\bar{r}=1$.
Imposing the physical boundary condition that the perturbation should vanish at spatial infinity, $\phi_\omega \to 0$,
we obtain the following solution for the $\omega=0$  mode:
\be
\label{w0}
\phi_{\omega=0}=C_0 
\sin
\left(
\frac{\bar{Q}}{\bar{r}}
\right)  \ ,
\ee
where $C_0$ is an integration constant. At the surface of the conductor different boundary conditions can be chosen. 
The Dirichlet (D) or Neumann (N) boundary conditions,
\begin{equation}
\label{bcs}
\phi_0(r_{\rm s})=0 \ \ \ (D) \ ,
\qquad
\phi_0 '(r_{\rm s})=0 \ \ \ (N) \ ,
\end{equation}
lead to
\be
\label{bcs_quad}
\sin
\left(
\bar{Q}
\right)=0 \ \ \ (D) \ ,
\qquad 
\cos
\left(
\bar{Q}
\right)=0  \ \ \ (N) \ ;
\ee
consequently, these boundary conditions yield, respectively, the following constraints on the parameters:
\be
\label{quad_cond}
\bar{Q}
\Big|_{D}
=\pi (1+n) \ ,
\qquad  
\bar{Q}
\Big|_{N}
=\frac{\pi}{2} (1+2n) \ , \qquad n\in \mathbb{N}_0 \ .
\ee
If  $\bar{Q}$ is greater than the  value given by \eqref{quad_cond} for $n=0$,
the lowest mode of $\phi_1$ has  nodes, implying there are eigenmodes with $\omega^2<0$. Then, the Coulomb solution is unstable.
Instability is therefore present if  the effective coupling $\bar{Q}$ is larger than a threshold value, which depends on the boundary condition chosen, i.e., when
\be
\label{critical}
\bar{Q}>\bar{Q}_D:=\pi \ ,
\qquad 
\bar{Q}>\bar{Q}_N:=\frac{\pi}{2} \ .
\ee

A third type of boundary condition can be defined, which will make contact with the numerical evolution in Section 
\ref{sec_num_sim}
\be 
(r\phi_{\rm \omega=0})'|_{r=r_{\rm s}}=0 \ .
\label{radbc}
\ee
This will be called a \textit{radiative} (R) boundary condition,
which corresponds to the ingoing boundary condition Eq.~(\ref{eq:ingoing condition}) for $\omega=0$.
From \eqref{w0} it leads to the condition
\be
\tan\left(
\bar{Q}
\right)
=
\bar{Q}
\ .
\ee
In this  case, the lowest mode gives $\bar{Q}=0$. 
Thus  for any value of $a,Q>0$, the Coulomb solution is \textit{always} unstable for these radiative boundary conditions.

\section{The linear model}
\label{sec3}
Let us now consider several specific choices of the coupling function $f(\phi)$. The first choice leads to an exact solution already discussed in~\cite{Herdeiro:2018wub}. 
Albeit not discussed explicitly therein, this choice linearizes the $\phi$ equation. Thus, we call it the \textit{linear} model. 

\subsection{The scalarized solution}
The linear model has the coupling function:
\be
\label{model1}
f(\phi)
&=&\frac{1}{1-a\phi^2} \ , \qquad \stackrel{\eqref{potmot}}{\Rightarrow} \qquad \mathcal{V}_{\rm eff}(\phi_0)=\frac{Q^2}{2}(a\phi_0^2-1) \ .
\ee
The effective potential $\mathcal{V}_{\rm eff}$ informs us that this coupling leads to a 1-dimensional mechanical model with a harmonic potential. 
It has  a minimum at  $\phi_0=0$;  this is not, however, the only solution to the equations of motion, as we shall promptly see. 
For  this model, the electric field \eqref{electric} reduces to
\be
A_0'
=\frac{Q}{r^2} (1-a \phi_0^2) \ .
\ee
One possible viewpoint is to face the scalarized solutions as introducing a sort of medium around the charged sphere. This interpretation has been used, say, in ~\cite{Herdeiro:2019iwl}. Then, the impact on the gauge field can be regarded  as a screening of the charge. The scalar field equation \eqref{xeq} then becomes the exactly linear (harmonic) equation
\be
\label{linear}
\frac{d^2\phi_0}{dx^2}=-aQ^2\phi_0 \ .
\ee
The solution satisfying the boundary condition $\phi(r\to \infty)=0$ is  given by, in terms of the $r$ variable,
\be
\label{quadratic}
&&\phi_0
=\zeta 
\sin
\left(
\frac{\sqrt{a}Q}{r}
\right) \ ,
\ee
where $\zeta$ is an undetermined amplitude. Observe  that this \textit{exact} solution for the scalar field in the  Ms model defined by the coupling~\eqref{model1}, precisely coincides with the solution for the $\omega=0$ \textit{linear perturbation} around the Coulomb solution in Eq.~\eqref{w0}.
In the limit of $r\to \infty$ Eq.~\eqref{quadratic} leads to
\be
\phi_0
=\frac{Q_s}{r}
+{\cal O}
\left(
\frac{1}{r^3}
\right),
\ee
where the scalar ``charge" $Q_s$~\footnote{Using common terminology, the coefficient of the leading asymptotic term of the scalar field, $Q_s$, is called ``charge"; but, in fact, it is not associated to any conservation law.} has been identified as
$Q_s
:=\sqrt{a}\zeta Q$.
Since the equation of motion~\eqref{linear} is linear,
$Q_s$ also grows linearly with the amplitude $\zeta$.

Imposing  
the boundary conditions (D) and (N) 
on the solution given by Eq.~\eqref{quadratic} at the surface of the conductor,
leads to, precisely, the conditions in Eq. \eqref{quad_cond}.
For each type of boundary conditions, there are, therefore, solutions with $n\in \mathbb{N}_0$ nodes between the conductor and infinity.
The lowest state, the $n=0$ node solution, branches off from the pure Coulomb solution at the threshold of instability against scalar perturbations; that is, when 
$\bar{Q}=\bar{Q}_N$ or $\bar{Q}=\bar{Q}_D$,
for the 
boundary conditions (N) or (D),
respectively,
given by Eq. \eqref{critical}. 
The branching is controlled by the amplitude $\zeta$; the electrostatic potential of the scalarized solution becomes~\cite{Herdeiro:2018wub}
\begin{equation}
A_0=-\left[\left(1-\frac{a\zeta ^2}{2}\right)\frac{Q}{r}
+\frac{\sqrt{a}\zeta ^2}{4}\sin \left(\frac{2\sqrt{a}Q}{r}\right)\right]\ .
\label{quadratic2}
\end{equation}
As we shall see in the next subsection, however, 
the scalarized solution Eqs.~\eqref{quadratic} and~\eqref{quadratic2} for the model \eqref{model1}
turns out to be unstable. 
This should be no surprise,
considering the effective potential~\eqref{model1}. 
Indeed, from~\eqref{xeq}, the (inverse) radial coordinate plays the role of ``time" in the 1-dimensional mechanical model. Thus the sinusoidal oscillations in the radial coordinate of solution~\eqref{quadratic} correspond to the time oscillations of a harmonic oscillator. Since such an oscillator has only a stable equilibrium point  at the minimum of the  potential, we anticipate that only the solution with $\phi_0=0$  everywhere could be stable. 
One can also relate this instability to the linear nature of~\eqref{linear}, which allows the amplitude of  the scalar field to
grow without bound. This is reminiscent of the instability observed for the scalarized black holes in extended-scalar-tensor-Gauss-Bonnet gravity, for a quadratic coupling function, that also leads to a scalar field equation which is linear in the scalar field~\cite{Silva:2017uqg,Doneva:2017bvd,Blazquez-Salcedo:2018jnn,Silva:2018qhn,Minamitsuji:2018xde,Blazquez-Salcedo:2020rhf,Blazquez-Salcedo:2020caw}.

\subsection{Instability of the scalarized charged conductor in the linear model}

To assess the linear stability of the scalarized solution given by Eqs.~\eqref{quadratic} and~\eqref{quadratic2}, we consider small radial perturbations about it:
\be
\label{radial_perturbations}
A_t=A_{0} (r)+ \epsilon a_t(t,r) \ , \qquad A_r= \epsilon a_r(t,r) \ , \qquad  \phi= \phi_0(r)+\epsilon \phi_1(t,r) \ ,
\ee
where $\epsilon\ll 1$ is the benchmark parameter  for the perturbations.
The details of the perturbation analysis  are shown in Appendix \ref{appendix1}.
By integrating out the gauge field perturbations $a_t$ and $a_r$,
one can obtain the master equation for the scalar field perturbation.
Fourier decomposing the master variable $\Phi$, 
$\Phi=\int d\omega \Phi_\omega e^{-i\omega t}$,
each mode obeys a time independent Schr\"odinger-type equation
\be
-\Phi_\omega''
+ V_{\rm eff}(r)
\Phi_\omega
=
\omega^2 \Phi_\omega \ ,
\ee
where we have defined the \textit{effective potential for the perturbations}
\be
\label{eff_potential}
 V_{\rm eff}(r)
:= 
-
\frac{Q^2}
         {2r^4 f(\phi_0(r))^3}
\left[
f (\phi_0(r))f_{\phi\phi}(\phi_0(r))
-2f_\phi(\phi_0(r))^2
\right]\ ,
\ee
with $f_{\phi\phi} (\phi):= d^2f(\phi)/d\phi^2$.
For the model \eqref{model1}, 
this effective potential simplifies to 
\be
V_{\rm eff}(r)
=-\frac{Q^2a}{r^4} \  ,
\label{potlin}
\ee
which is negative definite. This implies that unstable modes ($\omega^2<0$, see Appendix \ref{appendix2}) may exist.  In fact, they will exist if the perturbation with $\omega=0$ has nodes. The analysis then reduces to that of Sec.~\ref{sec2b}, and the  conclusion is that the scalarized charged conductors are unstable in the linear model  whenever the Coulomb solution, with the same charge and  the same boundary conditions, is unstable. Thus,  the scalarized solution cannot  be the endpoint of the instabiity of the Coulomb solution.

It is interesting to notice, in fact, that allowing for both a radial \textit{and} a time dependence in the ansatz~\eqref{san}, the model~\eqref{model1} admits the \textit{exact} time dependent solution
\be
\phi_0
=\zeta t
\sin
\left(
\frac{\sqrt{a}Q}{r}
\right) \ , \qquad A_0=-\left[\left(1-\frac{a\zeta ^2t^2}{2}\right)\frac{Q}{r}
+\frac{\sqrt{a}\zeta ^2t^2}{4}\sin \left(\frac{2\sqrt{a}Q}{r}\right)\right]\ ,
\ee
which simply amounts to giving the amplitude $\zeta$ a linear time dependence $\zeta \rightarrow \zeta t$. The existence of such an unbound (secular), growing solution, is again, related to the linear nature of the model. 

To obtain a physically self-consistent model,
in the next Section \ref{sec4} we shall consider  scalarized charged conductor solutions
for more general coupling functions,
wherein the nonlinear corrections in the scalar field equation
are able to quench the tachyonic instability.

\section{Nonlinear models}
\label{sec4}

We shall now consider two examples of coupling functions leading to models where the scalar equation of motion is nonlinear. As we shall see, in such cases the  scalarized solutions may be free of instabilities.

\subsection{The inverse quartic polynomial model}
The first model is a nonlinear continuation of the linear model~\eqref{model1}. Specifically, we consider model~\eqref{model1} augmented by a quartic term:
\be
\label{model2}
f(\phi)
&=&\frac{1}
{1-a\phi^2+
k^2a^2 \phi^4/4} \, \qquad \stackrel{\eqref{potmot}}{\Rightarrow} \qquad \mathcal{V}_{\rm eff}(\phi_0)=\frac{Q^2}{2}\left(-
\frac{k^2a^2 }{4}\phi_0^4+a\phi_0^2-1\right) \ .
\ee
The effective potential has extrema at
\be
\phi_0=0 \ \ ({\rm local \ minimum}) \ , \qquad \phi_0=\pm\sqrt{\frac{2}{ak^2}}   \ \ ({\rm maxima})\ .
\label{extrema_qurt}
\ee
With the choice~\eqref{model2}, the electric field \eqref{electric} becomes  
\be
A_0'
=\frac{Q}{r^2} 
\left(1-a\phi_0^2
+\frac{a^2k^2}{4}
\phi_0^4
\right).
\ee
Substituting this expression for the electric field into \eqref{seq},
the scalar equation becomes that of a nonlinear harmonic oscillator. Its solution, satisfying the boundary condition $\phi_0(r\to \infty)=0$ reads
\be
\label{quartic}
\bar\phi(\bar{r})
=
\sqrt{1-\sqrt{1-\bar C_0}
} \,
{\rm sn}
\left[
\frac{\bar Q}
     {\sqrt{2}\bar r}
\sqrt{1+\sqrt{1-\bar C_0}},
\frac{1-\sqrt{1-\bar C_0}}
     {1+\sqrt{1-\bar C_0}}
\right],
\ee
where ``${\rm sn}$" denotes a Jacobi elliptic function, 
\be
\label{redef}
\bar\phi:=\sqrt{\frac{ak^2}{2}}~\phi_0 \ \ \ \ (k\neq0) \ ,
\ee
$\bar{r}$ and $\bar{Q}$ were introduced in Eq. \eqref{barred_quantities},
and
$\bar C_0$ 
is an integration constant,
which will be fixed by the boundary conditions
at the surface of the conductor $r=r_{\rm s}$.  This form of the solution applies for $k\neq 0$; for $k=0$ 
the coupling~\eqref{model2} reduces to the coupling~\eqref{model1} and hence we recover the linear model.

\begin{figure}[h]
\unitlength=1.1mm
\begin{center}
 \includegraphics[height=7.5cm,angle=0]{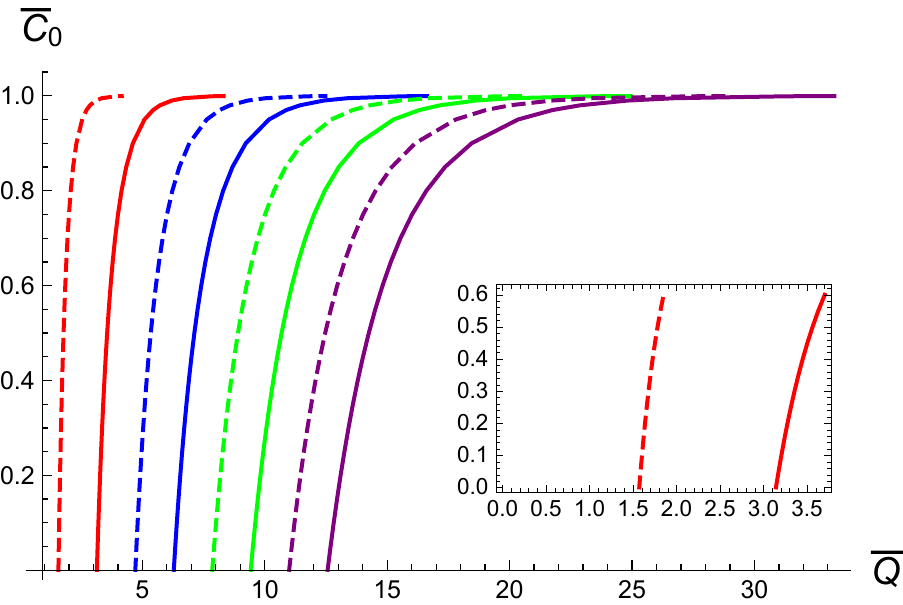} %
\caption{
$\bar C_0$ as a function of $\bar Q$ for the scalar field solution~\eqref{quartic}, in the model with an inverse quartic coupling~\eqref{model2}, and for
the boundary condition (D) (solid curves)  or 
the boundary condition (N) (dashed curves). 
Red, blue, green, and purple curves
correspond to solutions with
0-, 1-, 2- and 3-nodes, respectively.
Each curve is truncated at $\bar C_0=0.9995$.
(Inset) 
Zoom of the 0-node solutions around the origin.
}
  \label{a1}
\end{center}
\end{figure}

At first glance, the function \eqref{quartic} looks imaginary for $\bar{C}_0>1$.
However, we numerically confirmed that $\bar{\phi}$ remains real even 
if Eq. \eqref{quartic} is analytically continued into the parameter region of $\bar{C}_0>1$
and satisfies the boundary condition $\bar{\phi}(\bar{r}\to \infty)=0$, similarly to what holds for $0\leqslant\bar{C}_0\leqslant 1$,
although we did not find an alternative analytic form of \eqref{quartic} in which the arguments are explicitly real. In this  sense the solution  \eqref{quartic} is valid for $\bar{C}_0>0$.
We exclude the case of ${\bar C}_0<0$, for which Eq. \eqref{quartic} becomes imaginary.
As shown in Fig. \ref{a1},
the boundary conditions (N) and (D) \eqref{bcs}
admit the solutions $0\leqslant \bar{C}_0\leqslant 1$ for any $\bar{Q}$.
On the other hand, the solutions adapted to the radiative
boundary condition used in the numerical simulations (see Section~\ref{sec_num_sim}) 
include the case $\bar{C}_0>1$. We can then resort to numerical methods to obtain the corresponding  profiles of the scalarized solution of Eq.~\eqref{seq}. 
We also remark that for $\bar C_0=0$ one recovers the Coulomb solution with $\bar \phi=0$; for $\bar C_0=1$ one obtains 
\be
\label{special}
\bar\phi\to
\tanh 
\left(
\frac{{\bar Q}}{\sqrt{2}\bar r}  
\right) \ .
\ee

The asymptotic ($\bar r\to \infty$) behaviour of~\eqref{quartic} gives 
\begin{eqnarray}
\bar\phi=
\frac{{\bar Q}_s}{\bar r}
+{\cal O}
 \left(\frac{1}{\bar r^{3}}\right) \ ;
\end{eqnarray}
then, considering~\eqref{redef},  the scalar ``charge" ${\bar Q}_s$ is given by 
\be
\bar{Q}_s
=\frac{\sqrt{\bar{C}_0}\bar{Q}}
     {\sqrt{2}}
\ .
\ee
Solution~\eqref{quartic} contains two independent integration constants,  $\bar C_0$ and  $\bar Q$. These are, however, related by the  boundary condition on the conductor, $\bar r =1$. 
In Fig. \ref{a1} this relation is exhibited: 
${\bar C}_0$ is shown as a  function of $\bar Q$.
The solid and dashed curves
correspond to 
the boundary conditions (D) and (N)
at the surface of the conductor,
respectively,
where each curve is truncated at $\bar C_0=0.9995$.
Red, blue, green, and purple curves
correspond to the solutions with
0-, 1-, 2- and 3-nodes, respectively.
The inset 
provides a magnification of the 0-node solutions
around the bifurcation point  from the Coulomb solution.
The critical couplings \eqref{critical}
given by  
$\bar Q_N= \pi/2 \approx 1.5708$
and 
$\bar Q_D=\pi \approx 3.1416$,
are in good agreement with the bifurcation points  of the 0-node solutions, given in the inset of the figure.

Fig. \ref{a1} shows that for
the 0-nodes solutions, 
$\bar C_0$ approaches unity very quickly 
as a function of $\bar Q$.
The limit $\bar{C}_0\to 0$ reproduces Eq.~\eqref{bcs_quad}
giving rise to the solutions \eqref{quad_cond}
which correspond to vertical straight lines starting from the bifurcation points
in Fig.~\ref{a1}.
%

\subsection{The inverse cosine coupling}
As a second nonlinear completion of the coupling~\eqref{model1} (which led to the linear model), 
we consider the inverse cosine coupling \eqref{model3},
\be
\label{model3}
f(\phi)
&=&
\frac{1}{\cos\left(\sqrt{2a} \phi \right)} \qquad \stackrel{\eqref{potmot}}{\Rightarrow} \qquad \mathcal{V}_{\rm eff}(\phi_0)=-\frac{Q^2}{2}\cos\left(\sqrt{2a} \phi_0 \right) \ .
\ee

\begin{figure}[h]
\unitlength=1.1mm
\begin{center}
 \includegraphics[height=7.0cm,angle=0]{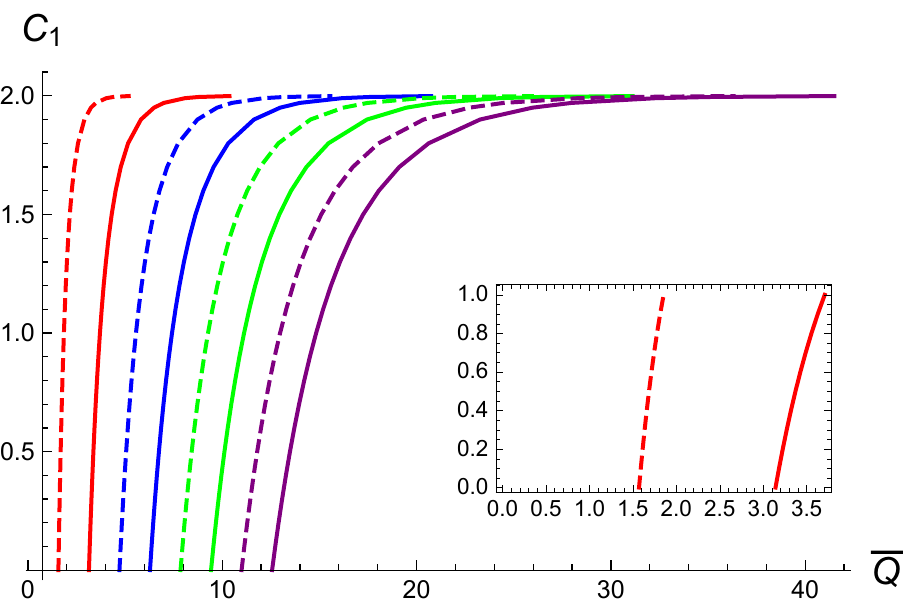} %
\caption{
$C_1$ as a function of $\bar Q$ for the scalar field solution~\eqref{cosine_sol}, in the model with an inverse quartic coupling~\eqref{model3}, and for 
the boundary condition (D) (solid curves) or the boundary condition (N) (dashed curves).
%
Red, blue, green, and purple curves
correspond to solutions with
0-, 1-, 2- and 3-nodes, respectively.
Each curve is truncated at $C_1=1.999$.
(Inset) 
Zoom of the 0-node solutions around the origin.
}
  \label{cosineQ}
\end{center}
\end{figure}

The effective potential has now  infinite countable set of extrema, at
\be
\sin\left(\sqrt{2a} \phi_0 \right)=0 \ .
\label{extrema3}
\ee
The electric field \eqref{electric} is now
\be
A_0'
=\frac{Q}{r^2}
 \cos \left( \sqrt{2a} \phi_0\right) \ .
\ee
The scalar field equation~\eqref{xeq} becomes  a 1-dimensional  sine-Gordon equation. Its solution, satisfying the boundary condition $\phi_0(r\to \infty)=0$,
can be expressed in terms of the Jacobi amplitude, denoted ``${\rm am}(k,m)$" as:
\be
\label{cosine_sol}
\bar\phi(r)
=
{\rm am}
\left(
\sqrt{\frac{C_1}{2}}
\frac{\bar Q}{\bar r},
\frac{2}{C_1}
\right)\ ,
\ee
where now
\be
\bar\phi:=\sqrt{\frac{a}{2}}~\phi_0 \ , 
\label{redef2}
\ee
and $C_1$ is an integration constant.
The function \eqref{cosine_sol} takes a real value
and satisfies the boundary condition $\bar{\phi}(\bar{r}\to \infty)=0$
for any $C_1>0$.
On the other hand,
we exclude the case of $C_1<0$, for which Eq. \eqref{cosine_sol} becomes imaginary.
In order to impose 
the (D) or (N)  boundary conditions \eqref{bcs},
we restrict to $0\leqslant C_1\leqslant 2$ (see Fig. \ref{cosineQ}).
In the lower limit, $C_1\to 0$, one verifies that
$\phi\to 0$, which represents the Coulomb solution.
In the upper limit, $C_1=2$, the solution becomes
\be
\bar\phi(r)=2\arctan^{-1}\left(e^{\bar Q/\bar r}\right)-\frac{\pi}{2}\ ,
\ee
which coincides with the 1-soliton solution of the sine--Gordon equation with no time dependence \cite{Rajaraman:1987}.

In the large distance limit,~\eqref{cosine_sol} yields
\begin{eqnarray}
\bar\phi= \frac{{\bar Q}_s}{\bar r}+ {\cal O} \left(\frac{1}{\bar r^{2}}\right) \ .
\end{eqnarray}
Thus, the scalar charge is 
$\bar{Q}_s:= \sqrt{C_1}\bar{Q}/\sqrt{2}$.

As in the previous subsection, $C_1$ is related to $\bar Q$ by imposing a boundary condition 
at the surface of the conductor 
$\bar{r}=1$.
This relation is exhibited in Fig.~\ref{cosineQ}, where
$C_1$ is shown as a function of $\bar Q$ for the 
boundary condition (D) (solid curves) and
the boundary condition (N) (dashed curves).
The red, blue, green, purple curves
correspond to the solutions of 
0-, 1-, 2- and 3-nodes, respectively.
Each curve is truncated at $C_1=1.999$.
The inset  
corresponds to the magnification of the 0-node solutions around the origin.
Thus, 
the behavior of this model is similar to the previous one, defined by \eqref{model2}.
The threshold coupling for which the Coulomb solution becomes unstable
is given by Eq. \eqref{critical}.

\subsection{Stability of the scalarized charged conductor in the nonlinear models}
In order for the scalarized solution to be the end state emerging from the evolution of the (unstable) Coulomb solution, it must be the preferred state. Let us first assess energetic preference, and then the dynamical one. 

The energy of the scalarized solution can be computed as:
\begin{align}
E&=4\pi\int_{r_{\rm s}}^\infty  {\rm d}r~r^2\left[\frac{1}{2}\phi'^2+\frac{1}{2}f(\phi)(A'_0)^2\right]
=2\pi\int_{r_{\rm s}}^\infty {\rm d}r~\frac{Q^2}{r^2}\left[C_i-1+\frac{2}{f(\phi(r))}\right] \nonumber \\
&=E_{\rm Coulomb}(C_i-1)+4\pi Q^2\int_{r_{\rm s}}^\infty{\rm d}r~\frac{1}{r^2f(\phi(r))} \ ,
\label{energy_ss}
\end{align} 
where $i=0, 1$, to account  simultaneously for each of the two models in the previous subsections, and $C_0:=\bar C_0/k^2$.

To illustrate which one is the \textit{energetically} preferred state, let us take concrete values for $C_i$.
If we choose $\bar C_0=1$, for the inverse quartic polynomial model, then we obtain
\be
E=E_{\rm Coulomb}-2\pi \bar Q^2\frac{r_{\rm s}}{a}\left\{1-\frac{2\sqrt{2}}{\bar Q}\tanh\left(\frac{\bar Q}{\sqrt{2}}\right)\left(1-\frac{1}{3}\tanh^2\left(\frac{\bar Q}{\sqrt{2}}\right)\right)\right\} \ .
\label{ene1}
\ee
For the solutions satisfying $\bar Q/\sqrt{2}>1.2964$, the second term is negative, so that $E<E_{\rm Coulomb}$.
If we choose $C_1=2$, for the inverse cosine model, we obtain
\be
E=E_{\rm Coulomb}-4\pi \bar Q\frac{r_{\rm s}}{a}\left(\frac{4}{1+e^{2\bar Q}}+\bar Q-2\right).
\ee
If $\bar Q>1.9150$, the second term is, again, negative.
Thus, this illustrates that, at least in some part of the parameter space, the scalarized solutions have less energy than the Coulomb solution.

Next, we consider the linear stability of the scalarized solutions to assess if they can be, dynamically, the end state.
For the models \eqref{model2} and \eqref{model3}
with the nonlinear corrections in the scalar field equations of motion,
 introducing the dimensionless quantities \eqref{barred_quantities} 
the effective potential for the radial perturbations \eqref{eff_potential}
is given by,  respectively,
\be
\label{effpot2}
V_{\rm eff}(\bar r)
&=&
-\frac{\bar Q^2}{\bar r^4} 
 \left(1-3\bar\phi_0(\bar r)^2\right) \ ,
\\
\label{effpot3}
V_{\rm eff}(\bar r)
&=&
-\frac{\bar Q^2}{\bar r^4} 
 \cos \left(2 \bar\phi_0(\bar r)\right) \ .
\ee
This informs us  that when  the amplitude of the background scalar field $\bar\phi_0(r)$ becomes sufficiently large,
the effective potential becomes non-negative in part of the spacetime region, unlike that for the linear model~\eqref{potlin}, which is everywhere  negative. We can therefore understand the (potential) healing effect of the nonlinear terms, which may indeed 
 quench the instability.

\begin{figure}[h]
\unitlength=1.1mm
\begin{center}
 \includegraphics[height=5.1cm,angle=0]{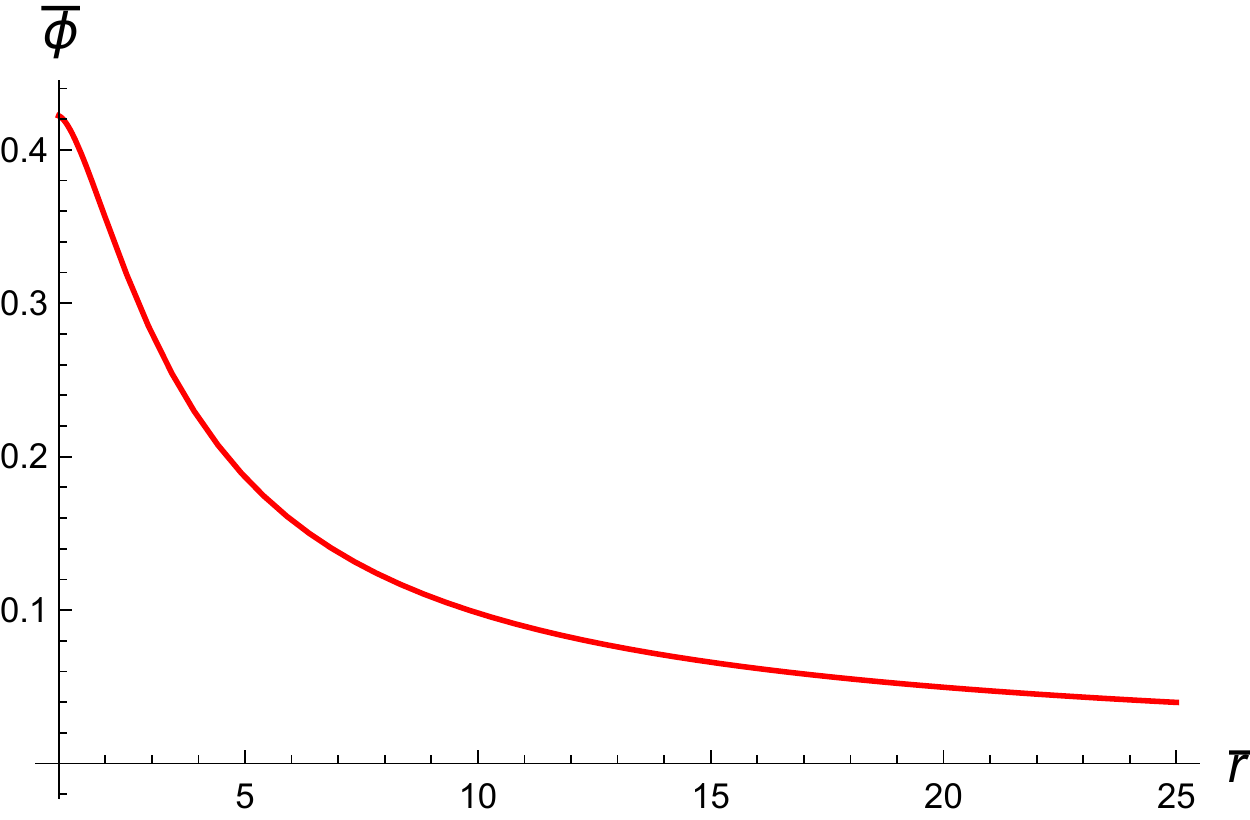} %
 \includegraphics[height=5.1cm,angle=0]{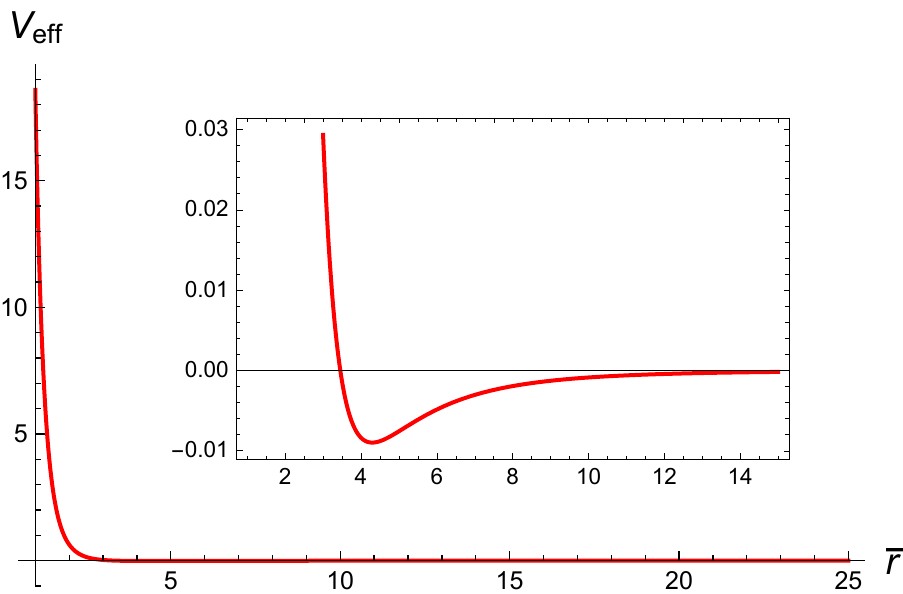} %
\caption{
$\bar\phi_0$ (left panel) and $V_{\rm eff}$ (right panel) as functions of $\bar r$ for the model \eqref{model2} 
and 
the boundary condition (N).
We set  $\bar Q=3.22738$ and $\bar{C}_0=0.995$. 
The inset shows  a zoom of $V_{\rm eff}$, to clearly exhibit the negative region.
}
  \label{veff2}
\end{center}
\end{figure}

\begin{figure}[h]
\unitlength=1.1mm
\begin{center}
 \includegraphics[height=5.1cm,angle=0]{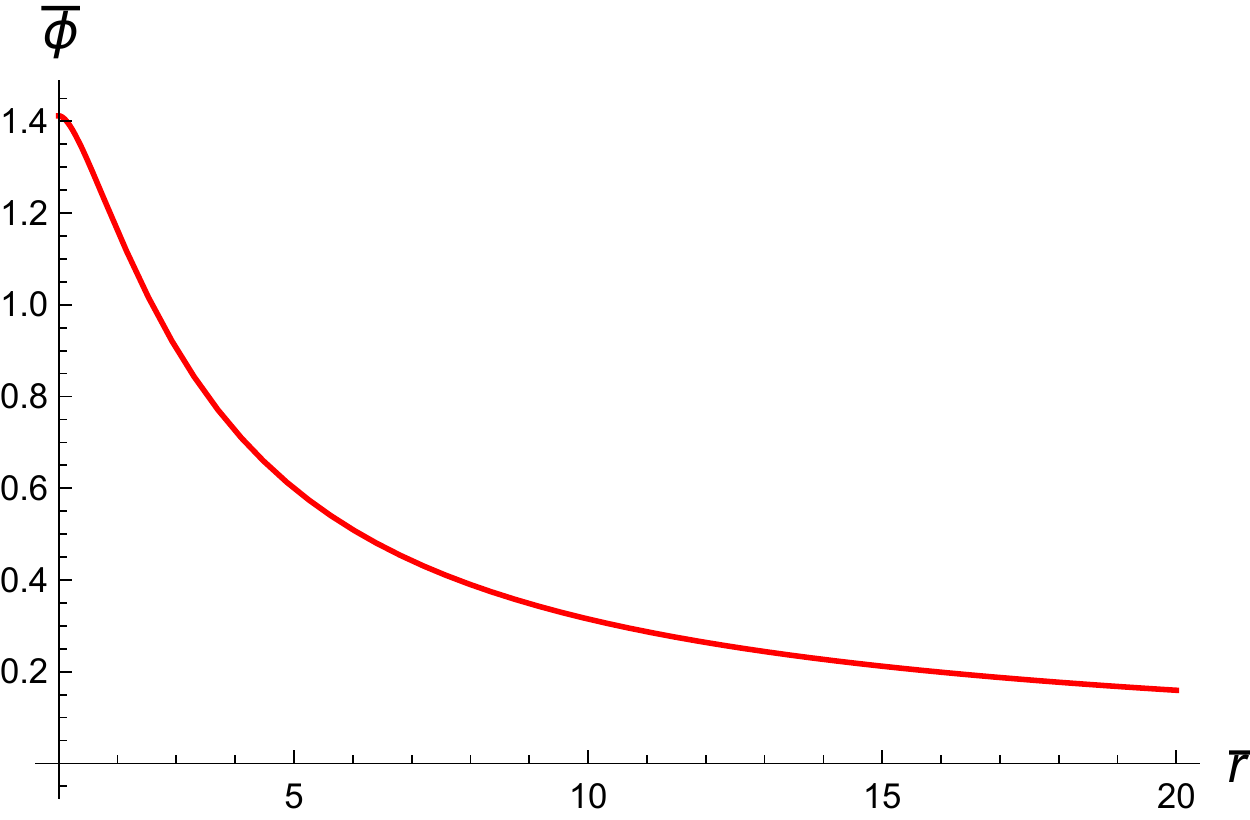} %
 \includegraphics[height=5.1cm,angle=0]{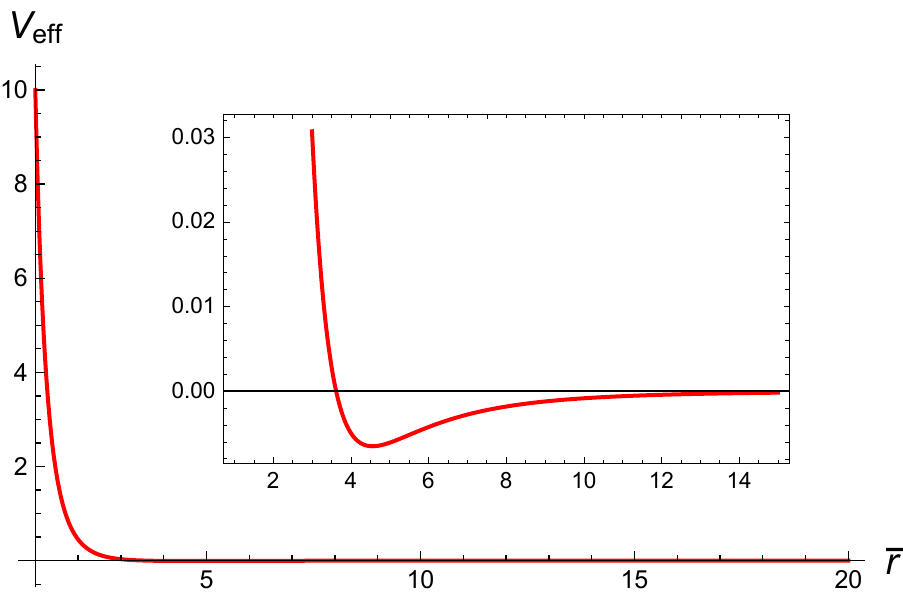} %
\caption{
Same as Fig.~\ref{veff2} but now for the model \eqref{model3} and 
the boundary condition (N).
We set  $\bar Q=3.24484$ 
 and $C_1=1.95$.
}
  \label{veff3}
\end{center}
\end{figure}

Next, we investigate if for concrete solutions the above effective potential indeed becomes positive. In Figs.~\ref{veff2} and~\ref{veff3}
we show the plots for the effective potential for radial perturbations
for the models \eqref{model2} and \eqref{model3}; 
$\bar\phi_0$ and $V_{\rm eff}$ are shown as functions of $\bar r$
for illustrations of scalarized solutions satisfying %
the boundary condition (N).


Figs.~\ref{veff2} and \ref{veff3} confirm that, for both the models,
the effective potential becomes positive in the vicinity of the  conductor
and hence 
there is no tachyonic mode trapped
in the vicinity of $r=r_{\rm s}$. This confirms the healing potential of the nonlinear models, concerning the instability of the scalarized solution. However,
the effective potential is still (mildly) negative 
at larger radius (see the insets in Figs.~\ref{veff2} and \ref{veff3}).\footnote{This is unlike what happens for black hole scalarization, see e.g.~\cite{Fernandes:2019rez}.}
Thus, we must further analyze the perturbation equation in order to establish the  stability of the scalarized solutions. 

By the same procedure that was followed for the linear model, introducing the dimensionless quantities \eqref{barred_quantities},
we have obtained the scalar field perturbation equation around the scalarized solution in a Schr\"odinger-like form:
\be
\label{dimless_Sch}
-\Phi''_\omega(\bar r)+V_{\rm eff}\Phi_\omega=\bar{\omega}^2\Phi_\omega \ ,
\ee
where now the prime denotes the derivative 
with respect to $\bar r$ and $V_{\rm eff}$ is given by Eq.~\eqref{effpot2} or Eq.~\eqref{effpot3},
and we have introduced $\bar{\omega}:=\omega r_s$.

To solve~\eqref{dimless_Sch} we need  to impose appropriate boundary conditions for the perturbations. A first possibility are standard quasinormal mode-like boundary conditions: we impose outgoing (ingoing) boundary conditions at infinity (on the conductor's surface), for the perturbations:
\begin{align}
\left.\left(\Phi'_\omega-i\bar{\omega}\Phi_\omega\right)\right|_{\bar r=\infty}&=0 \ , &&{\rm outgoing~condition},\\
\left.\left(\Phi'_\omega+i\bar{\omega}\Phi_\omega\right)\right|_{\bar r=1}&=0 \ , &&{\rm ingoing~condition}.\label{eq:ingoing condition}
\end{align}
This generalizes the radiative boundary conditions introduced before for $\omega=0$, cf. Eq.~\eqref{radbc}, since $\Phi:= r\phi_1$, cf. Appendix \ref{appendix1}, and justifies the terminology ``radiative" boundary condition.
As two other possibilities, we impose a Dirichlet (Neumann) boundary condition at the conductor's surface instead of an ingoing boundary condition,
which is equivalent to the boundary conditions (D), (N) used for obtaining the static solution and the numerical simulations in Sec.~\ref{sec_num_sim}
and hence useful for interpreting the result of the latter,
\begin{align}
\left.\left(\Phi'_\omega-i\bar{\omega}\Phi_\omega\right)\right|_{\bar r=\infty}&=0 \ , &&{\rm outgoing~condition},\\
\left.\Phi_\omega\right|_{\bar r=1}&=0 \ , &&{\rm Dirichlet~condition}.\label{eq:dirichlet condition}
\end{align}
\begin{align}
\left.\left(\Phi'_\omega-i\bar{\omega}\Phi_\omega\right)\right|_{\bar r=\infty}&=0 \ , &&{\rm outgoing~condition},\\
\left.\left(\frac{\Phi_\omega}{r}\right)'\right|_{\bar r=1}&=0 \ , &&{\rm Neumann~condition}.\label{eq:neumann condition}
\end{align}

\begin{figure}[htbp]
\centering
\includegraphics[width=7cm]{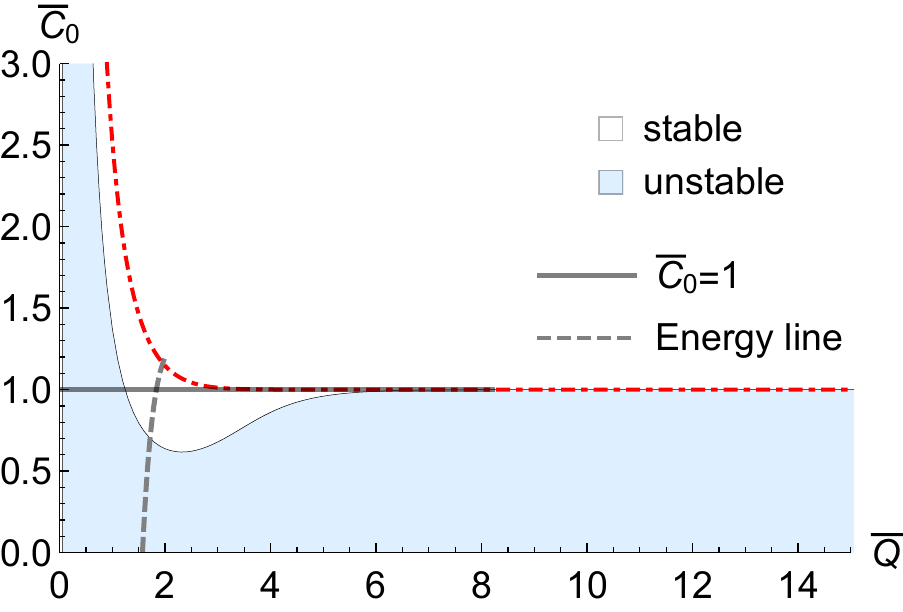}~
\includegraphics[width=10cm]{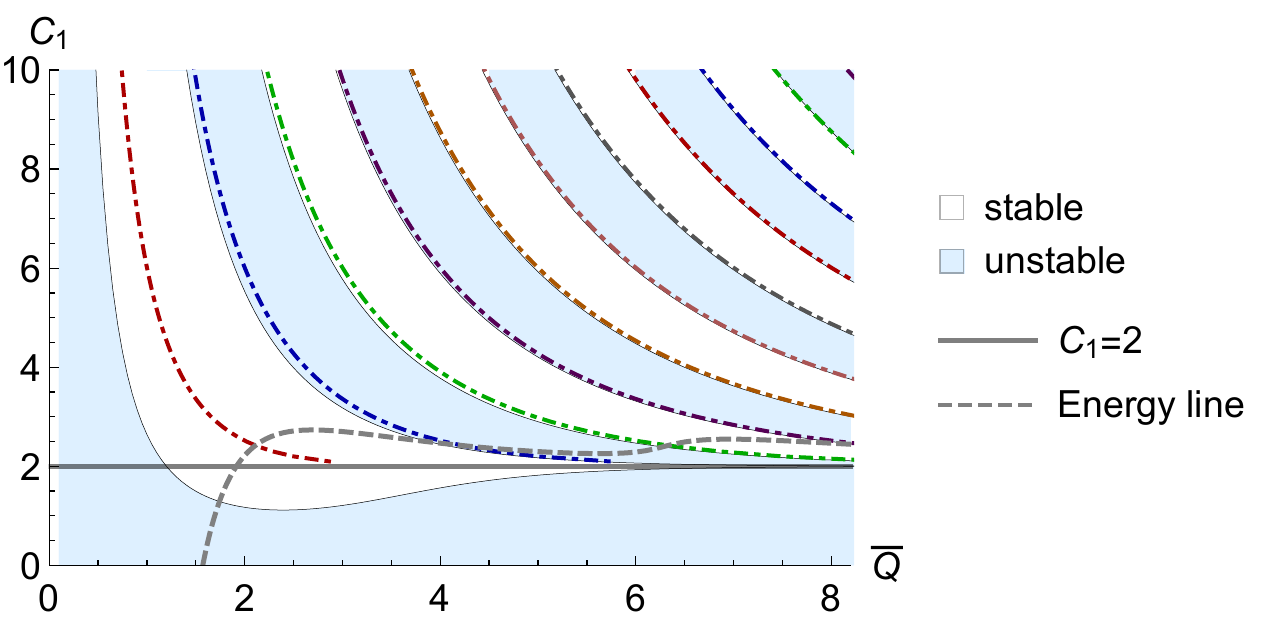}\\
\hspace{-3cm}
\includegraphics[width=7cm]{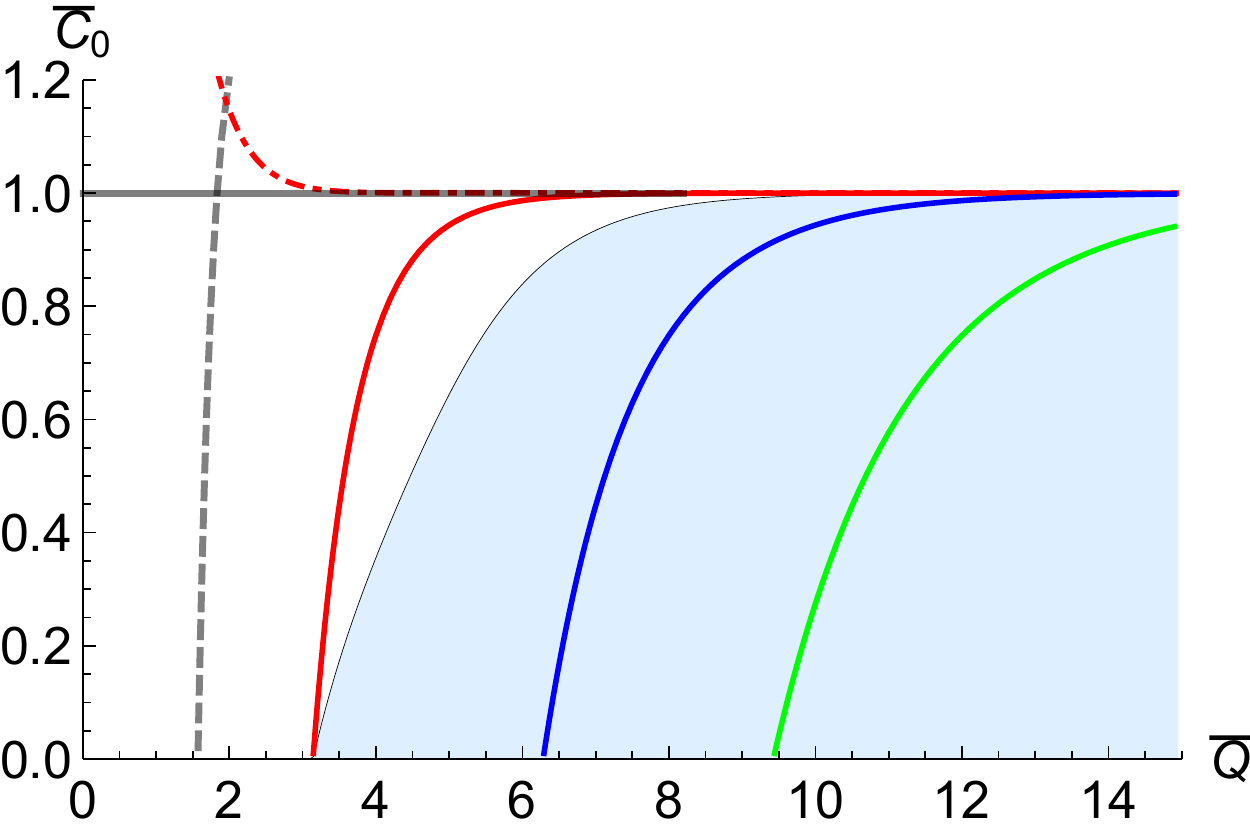}~
\includegraphics[width=7cm]{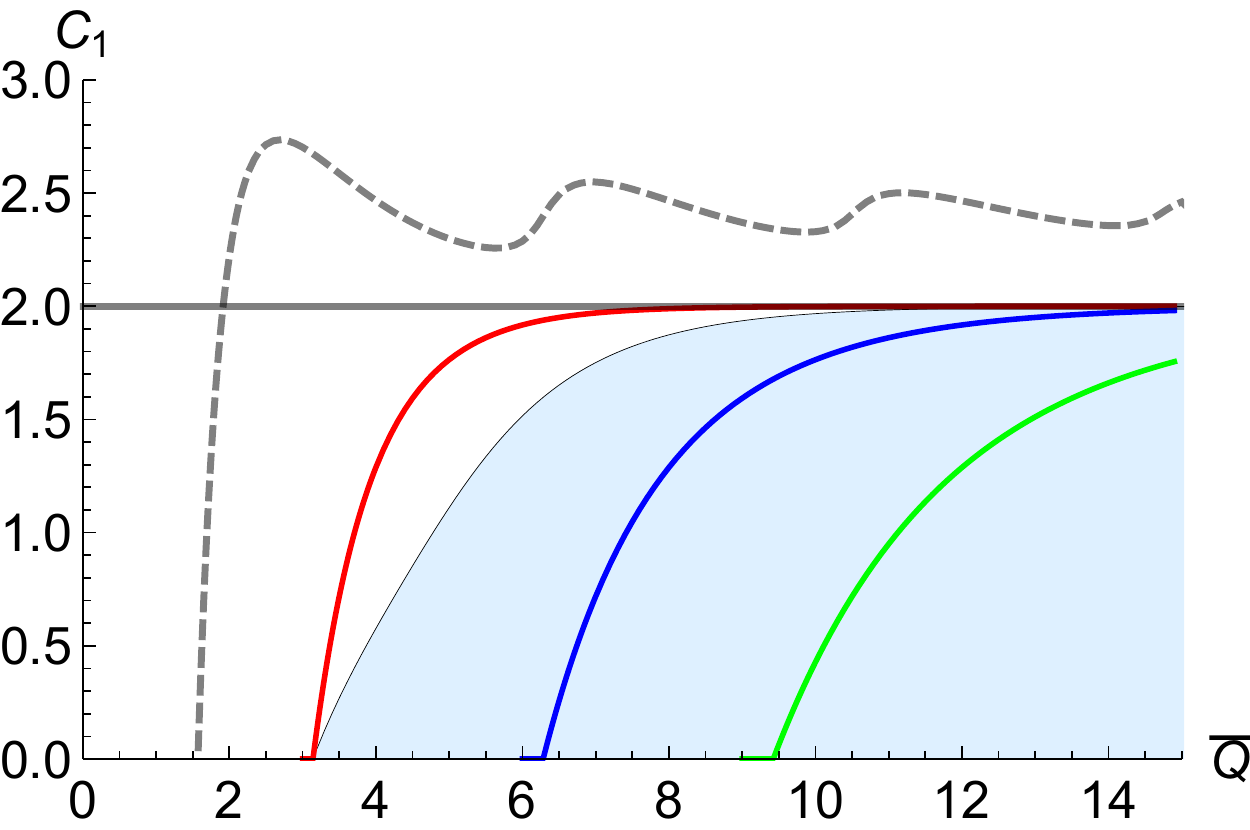}\\
\hspace{-3cm}
\includegraphics[width=7cm]{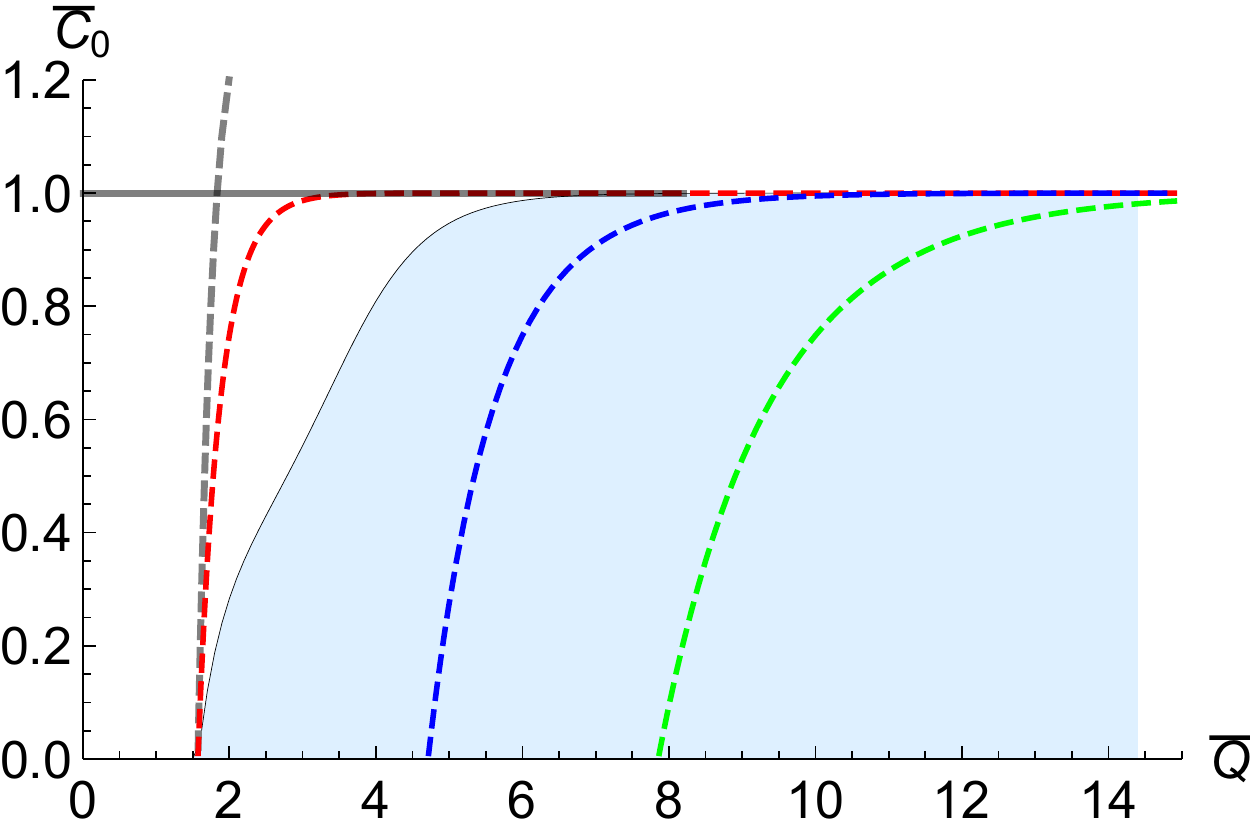}~
\includegraphics[width=7cm]{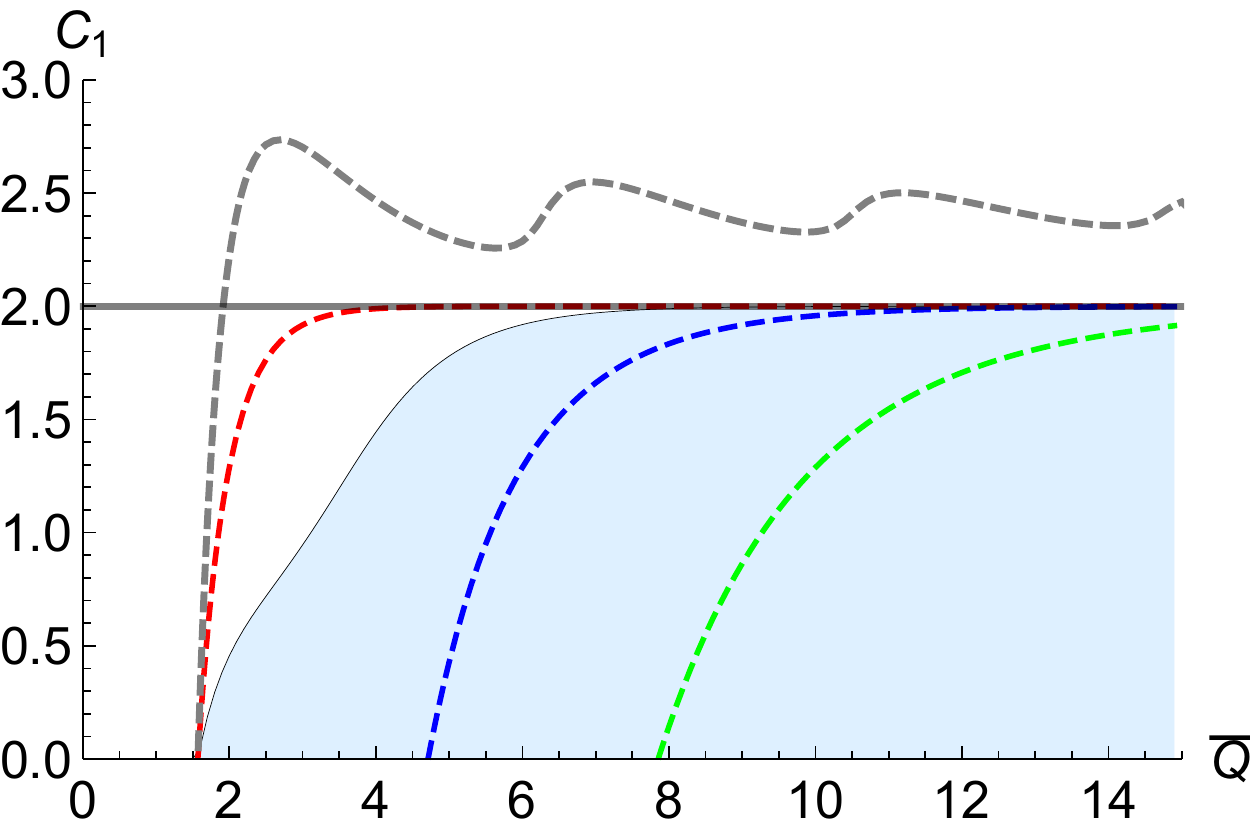}
\caption{(Un)Stable region of the scalarized solution in the two considered nonlinear models: Eq.~\eqref{quartic} (left panels) and Eq.~\eqref{cosine_sol} (right panels).
The boundary condition for the perturbation, at the conductor, is taken as the ingoing condition (top panels), 
Dirichlet condition (middle panels), and Neumann condition (bottom panels).
The gray dashed lines (``Energy line") correspond to the critical value of $\bar Q$ above which the scalarized solution has less energy than the Coulomb solution.
The curved  lines correspond to the static solutions satisfying  the boundary conditions (D) (solid curves),  or  (N) (dashed curves) or (R) [also enforcing a potential minimum
Eqs.~\eqref{extrema_qurt} and \eqref{extrema3} at the surface $\bar r=1$]  (dotdashed curves). The correspondence between the line color and the node of the solution is the same as Figs.~\ref{a1} and \ref{cosineQ} for 
the boundary condition (D) and the boundary condition (N). 
For the potential minimum condition, we can write the solution of Eq.~\eqref{extrema3} as $\bar\phi=n\pi/2$, where $n$ is an integer, and red, blue, green, purple... lines correspond to the solution with $n=1, 2, 3...$. }
\label{unstable}
\centering
\end{figure}

In order to obtain the parameter space regions of (in)stability, 
we follow the method  already used in Section~\ref{sec2b}.
Focusing on the static solutions (for the perturbations, $\bar{\omega}=0$) and confirming 
that the solution does not cross the horizontal axis,
we obtain the stability regions in Fig.~\ref{unstable}. The instability regions are highlighted as  the shaded blue regions in the  different panels. 
In this analysis we have left the constants $\bar C_0$ and $C_1$ as arbitrary (in terms of $\bar Q$),
in order to accommodate an arbitrary boundary condition for the scalarized solution. Then,  to identify the (background) scalarized solutions with the different boundary conditions, in the different plots we added solid, dashed, and dotdashed lines,  corresponding to the scalarized solutions  with boundary conditions (D),  (N)  and (R)\footnote{The last ones require the field value that
extremizes the effective potential to hold at the conductor; they correspond to the radiative boundary condition used in the numerical evolution of the next section.}, respectively. The  colours of these lines define the  number of nodes of the solution,  being the same as in Figs.~\ref{a1} and~\ref{cosineQ}  for (D) and (N) boundary conditions; in  particular red and blue colors correspond to 0 and 1 node solutions,  respectively.

Fig.~\ref{unstable} (left panels) exhibits the behavior for the inverse quartic polynomial model and ingoing/radiative boundary conditions (top), (D) boundary conditions (middle), and (N) boundary conditions (bottom) for the  perturbations. 
Each panel also shows the static scalarized solutions with corresponding boundary conditions (with 0, 1, or 2 nodes) respectively.
The gray dashed line shows the result obtained from Eq.~\eqref{energy_ss} about energetic preference: scalarized  solutions  to the right (left)  have less (more) energy than the  Coulomb solution  with the same charge. The following conclusions can be  drawn from these three panels:
\begin{itemize}
\item[i)] (Top panel) The (in)stability boundary line crosses $\bar C_0=1$ at some value of $\bar Q$, below which all scalarized solutions with $\bar C_0\leqslant 1$ are unstable.
This may be interpreted from the fact that the effective potential approaches the Coulomb solution's one for small charge or coupling and,  as we have seen, the Coulomb solution  is always  unstable against radiative boundary conditions.
On the other hand, the boundary line approaches $\bar C_0=1$ at large charge or coupling.
\item[ii)] (Middle and bottom panels)  There is no unstable region above $\bar{C}_0=1$.
Since the limit $\bar{C}_0\rightarrow0$ yields the Coulomb solution, 
the boundary line of the stability  reproduces the previous analysis of the stability of the Coulomb solution:
 along the $\bar{Q}$ axis, instability holds 
for $\bar{Q}>\bar{Q}_D=\pi$ (middle) and $\bar{Q}>\bar{Q}_N=\pi/2$ (bottom)
(see Eq. \eqref{critical}).
\item[iii)] (All panels) Only the 0-nodes scalarized solutions fit into the stable  areas. This fits the generic expectation that 0-node solutions are fundamental states and non-zero node solutions are excited states. Moreover, there is no implication (either way) between energy preference and dynamical stability.
\end{itemize}

Fig.~\ref{unstable} (right panels) exhibits the behavior for the inverse cosine model and ingoing/radiative boundary conditions (top), (D) boundary conditions (middle),
and (N) boundary conditions (bottom) for the  perturbations. 
Again, some observations can be made:
\begin{itemize}
\item[i)] (Top panel)  Stable regions now alternate  with unstable regions, due to the oscillating behavior of the effective potential, Eq.~\eqref{model3}.
  Now, the raditative boundary condition [imposing also that extermizes the effective potential~\eqref{model3} and thus obey~\eqref{extrema3} (equivalently, $\sin(2\bar\phi_0)=0$) at the conductor, a requirement that will become clear in the next section] leads to $2\bar\phi_0=n\pi$, where $n$ is an integer.
We can see that odd (even) $n$ solutions lie in the stable (unstable) region.
The energy preference line (the gray dashed line) behaves in a similar manner to the previous model for small $\bar Q$.
However, due to the existence of unstable regions, the line roughly follows the boundary lines of the stability.
\item[ii)] (Middle and bottom panels)  
There is no unstable region above  $C_1=2$ and only 0-nodes solutions are stable.
Again,  the limit $C_1\rightarrow0$ yields the Coulomb solution; as   such along the $\bar{Q}$ axis 
instability occurs for $\bar{Q}>\bar{Q}_D=\pi (\bar{Q}_N=\pi/2)$
(see Eq. \eqref{critical}).
\end{itemize}

To conclude: scalarized solutions with 0-nodes are possible final states  of the scalarization instability of the Coulomb solution,  in some ranges  of the parameter space and for the different boundary conditions. The next section will confirm this expectation, via fully dynamical numerical evolutions.

\section{Time evolutions}
\label{sec_num_sim}
In the previous sections we have constructed scalarized conducting spheres in Ms models, under specific choices of the nonminimal coupling function $f(\phi)$, namely Eqs.~\eqref{model1}, ~\eqref{model2}, and~\eqref{model3}. Then, we have investigated the linear stability of these solutions. For the nonlinear models~\eqref{model2} and~\eqref{model3}, this analysis suggests that there are stable scalarized solutions in appropriate parameter regions, which could be the end state of the evolution of the unstable Coulomb conducting sphere, when embedded in the corresponding Ms model. In order to establish if such scalarized solutions are indeed the final state of the dynamics,  in this section we numerically solve the time evolution of the Ms system.  
This will also allow us to investigate \textit{how} the final state is reached.

Performing time evolution implies dropping the stationarity assumption, which has been used in the previous sections to compute the scalarized solutions. Nonetheless, even dropping this assumption, the only relevant component of the electromagnetic field remains the radial one, $E^{r}=F^{tr}$, due to  spherical and gauge symmetries.
Therefore, the expression for the electromagnetic field is the same as in the static case and 
we obtain the scalar field equation from Eqs.~\eqref{eq1} and \eqref{eq2} as follows:
\begin{eqnarray}
-\ddot\phi(t, r)+\phi''(t, r)+\frac{2}{r}\phi'(t, r)=-\frac{Q^{2}f_\phi(\phi)}{2r^{4}f(\phi)^{2}} \ .
\label{Eq.KG eq.}
\end{eqnarray}

In the following,  we shall solve the time evolution of Eq.~(\ref{Eq.KG eq.}) numerically around the conductor.
To do so, let us define new variables $\varphi$,~$\pi$ as 
\be
\varphi:=r\phi \ , 
\qquad 
\pi:=r
\dot{\phi}\ .
\ee\
Using these variables, Eq.~(\ref{Eq.KG eq.}) is brought to  a first order system  of two equations:
\begin{eqnarray}
\dot{\varphi}
&=&\pi \ ,\\
\dot{\pi}
&=&
\varphi''
+\frac{Q^{2}f_\phi(\varphi/r)}{2r^3f(\varphi/r)^{2}} \ .
\end{eqnarray}
The conducting sphere's radius is $r_{\rm s}(> 0)$; thus, the physical domain within the radial direction is $[r_{\rm s},\infty[$.
The boundary condition imposed on the scalar field at the conductor must be chosen, as in the previous sections.
In the subsequent analysis, we consider three different kind of boundary conditions during the numerical evolution.
The first boundary condition is 
\begin{eqnarray}
\dot{\pi}&=&\pi'+\frac{f_{\phi}}{f^{2}}\frac{Q^{2}}{2r^{3}}\ , 
\label{radiative BC1} \\
\dot{\varphi}&=&\pi\,,
\label{radiative BC2}
\end{eqnarray}
on the surface of the conductor.
At the end state ($\dot\pi=\dot\varphi=0$),~\eqref{radiative BC1}-\eqref{radiative BC2} implies that, at the conductor, 
\begin{eqnarray}
\label{ext}
\frac{f_{\phi}}{f^2}=0 \ .
\end{eqnarray}
This boundary condition for the perturbation around the Coulomb solution is compatible with Eq.~(\ref{radbc})\footnote{
To check this point, let us consider perturbations around the Coulomb solution under the boundary condition  given by Eq.~(\ref{radiative BC2}).
The perturbations can be expressed as a sum of  an ingoing mode $\varphi_{+}(t+r)$ and an outgoing mode $\varphi_{-}(t-r)$.
From the boundary condition, we can show the outgoing mode vanishes on the conductor; thus 
the  boundary condition Eq. \eqref{radiative BC2} corresponds to Eq. \eqref{eq:ingoing condition}.
}
and hence it is radiative/outgoing. Moreover,~\eqref{ext} informs us that at the endpoint the scalar field
is at an extremum of the effective potential~\eqref{potmot} at the surface of the conductor. 
This makes direct contact with the stability analysis discussed in the top panels of Fig.~\ref{unstable}.
The second boundary condition is a Dirichlet (D) boundary condition:
\begin{eqnarray}
\dot{\pi}= 0\ , \qquad  
\dot{\varphi} = \pi\ ,
\label{Dirichlet BC}
\end{eqnarray}
on the surface of the conductor.
In this case, the scalar field value is always zero at the conductor.
Then, scalarized solutions with 
the boundary condition (D) are realized as the end states.  This makes direct contact with the stability analysis discussed in the bottom panels of Fig.~\ref{unstable}.
The third boundary condition is a Neumann (N) boundary condition:
\begin{eqnarray}
\partial_{r}\left(\frac{\varphi}{r}\right)&=&0\ ,\\
\partial_{r}\left(\frac{\pi}{r}  \right)&=&0\ ,
\label{Neumann BC}
\end{eqnarray}
on the surface of the conductor.
In this case, the radial derivative of the scalar field $\phi$ is always zero at the conductor,
and scalarized solutions with
the boundary condition (N) are realized as the end states.
In the asymptotic region, we impose an outgoing boundary condition.

To perform the evolution a \textsc{c++} time evolution code was written.
Time integration is calculated using a 4th order Runge-Kutta method, and spatial derivatives are evaluated using a 4th order central difference scheme.
Ghost zones (grid areas beyond inner boundary and outer boundary) for each boundary are evaluated using  a 2nd order Lagrange extrapolation. 
The numerical code has, overall, a 2nd order convergence.

The general behavior of the evolution does not strongly depend on the chosen initial data.
Here, we use momentarily static Gaussian initial data:
\begin{eqnarray}
\varphi(r,t=0)&=&Are^{-\left(
\frac{r - r_{0}}{w}
\right)^{2}}\ ,\\
\pi(r,t=0)&=&0 \ ,
\label{inidat}
\end{eqnarray}
where $A,r_{0}$, and $w$ are the amplitude, position, and width of the Gaussian pulse, respectively.
In the graph of the simulation, we use $\bar{\phi}=\sqrt{\frac{ak^{2}}{2}}\phi$ for the inverse quartic polynomial model,
and $\bar{\phi}=\sqrt{\frac{a}{2}}\phi$ for the inverse cosine model.

\subsection{Numerical results}
We shall now report the results obtained from the numerical simulations.
We shall use the scaled charge $\bar Q$, cf. Eq.~\eqref{barred_quantities}, in order to compare the results with the scalarized solutions obtained in the previous sections.

\subsubsection{Inverse quartic polynomial model}
\begin{figure}[htbp]
\centering
\includegraphics[width=7.8cm]{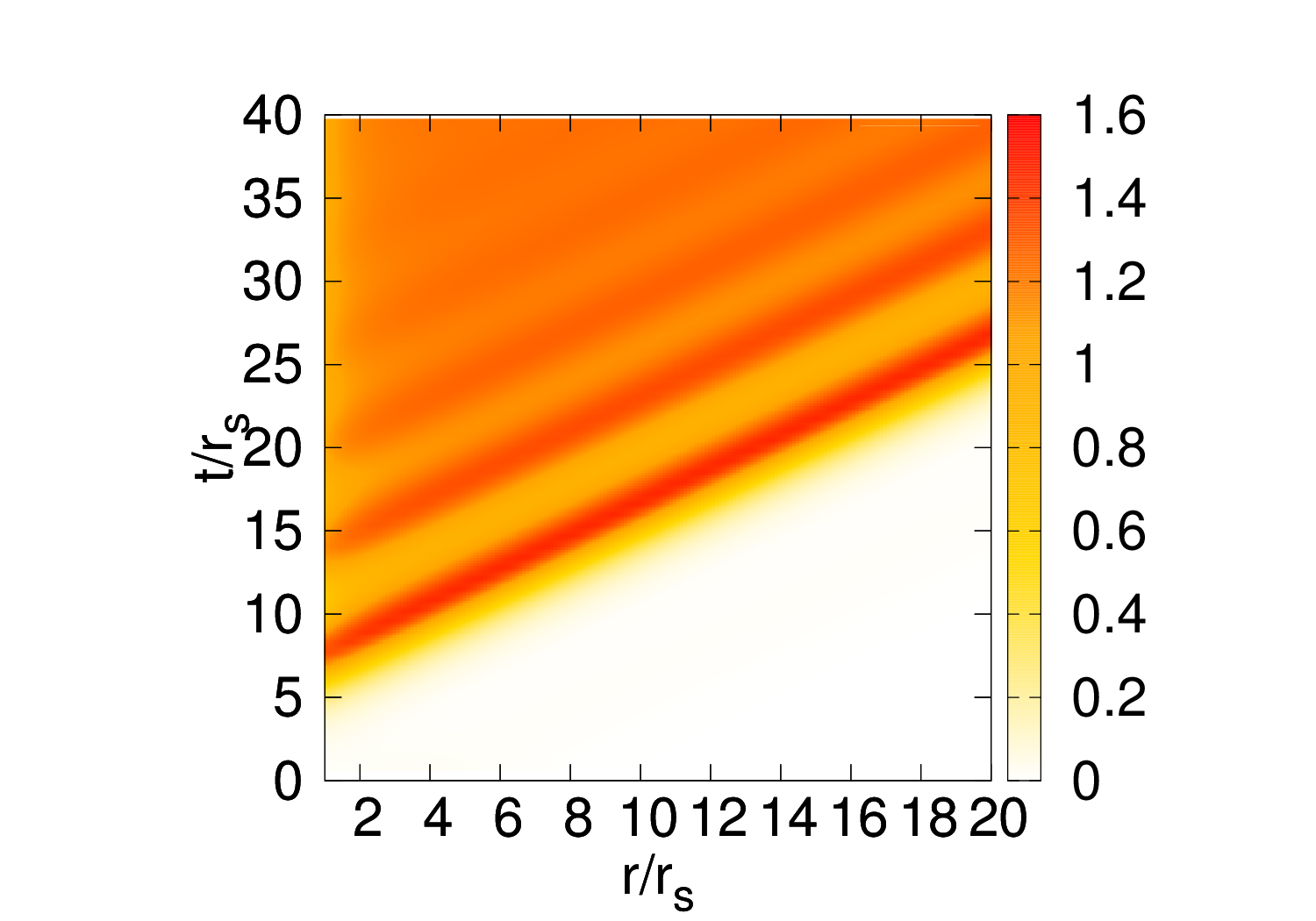}
\includegraphics[width=7cm]{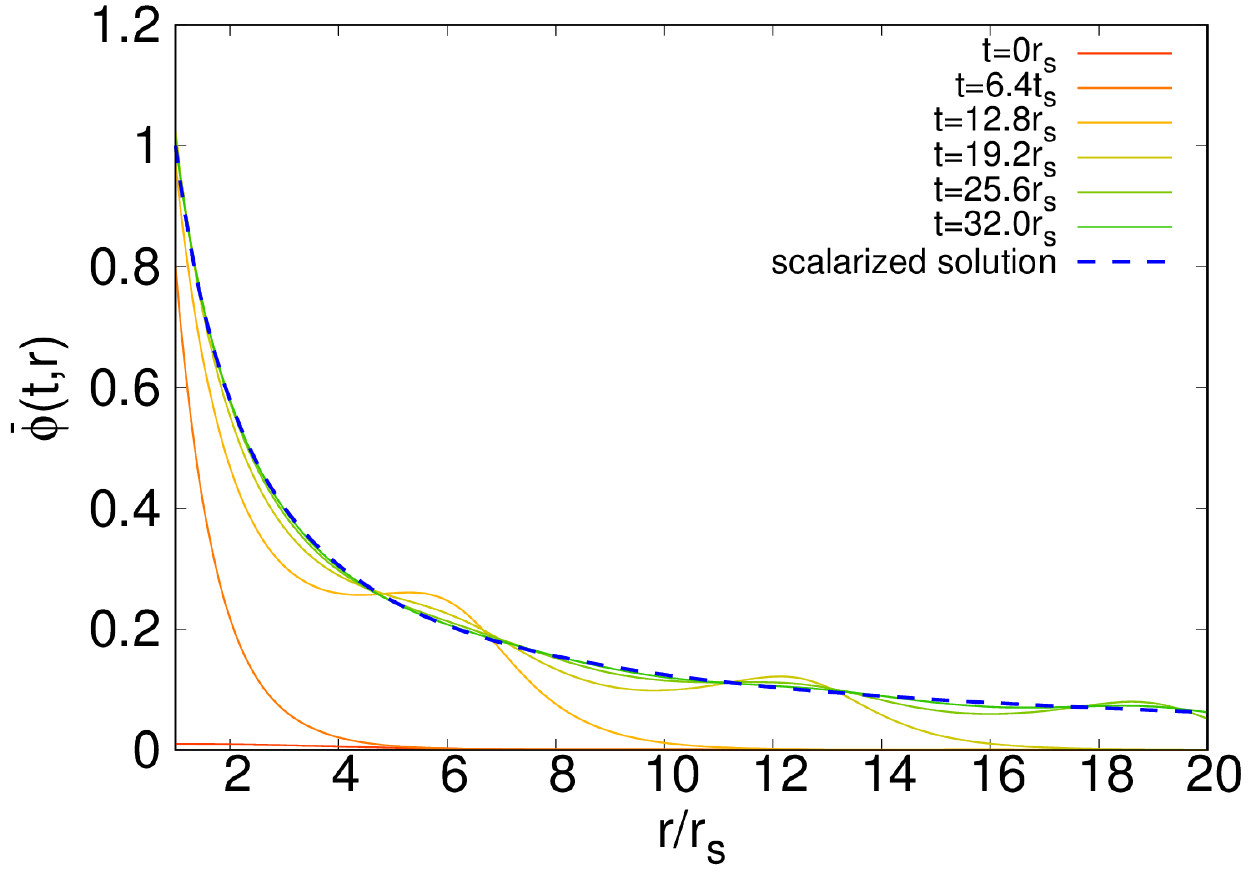}\\
\includegraphics[width=7.8cm]{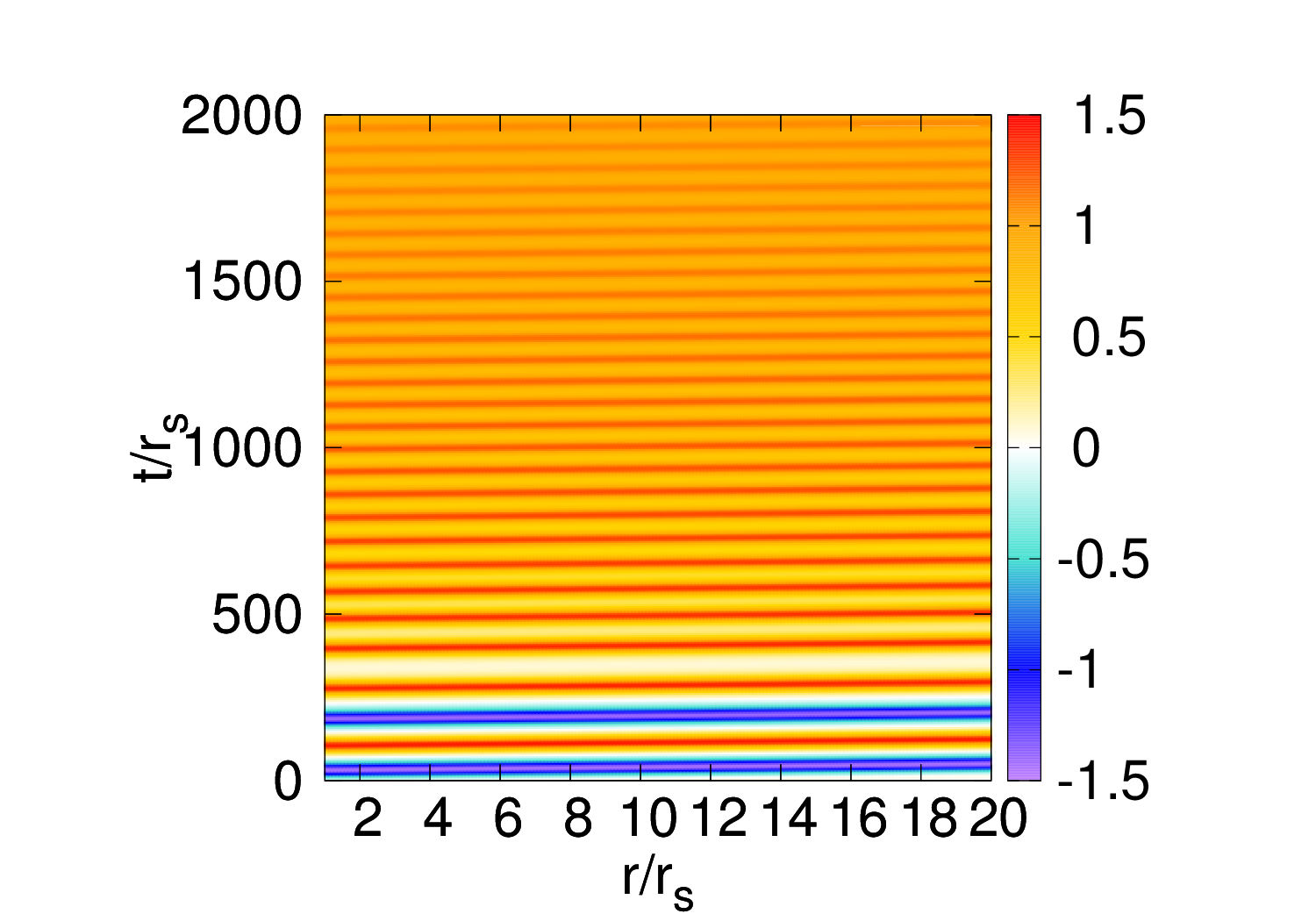}
\includegraphics[width=7cm]{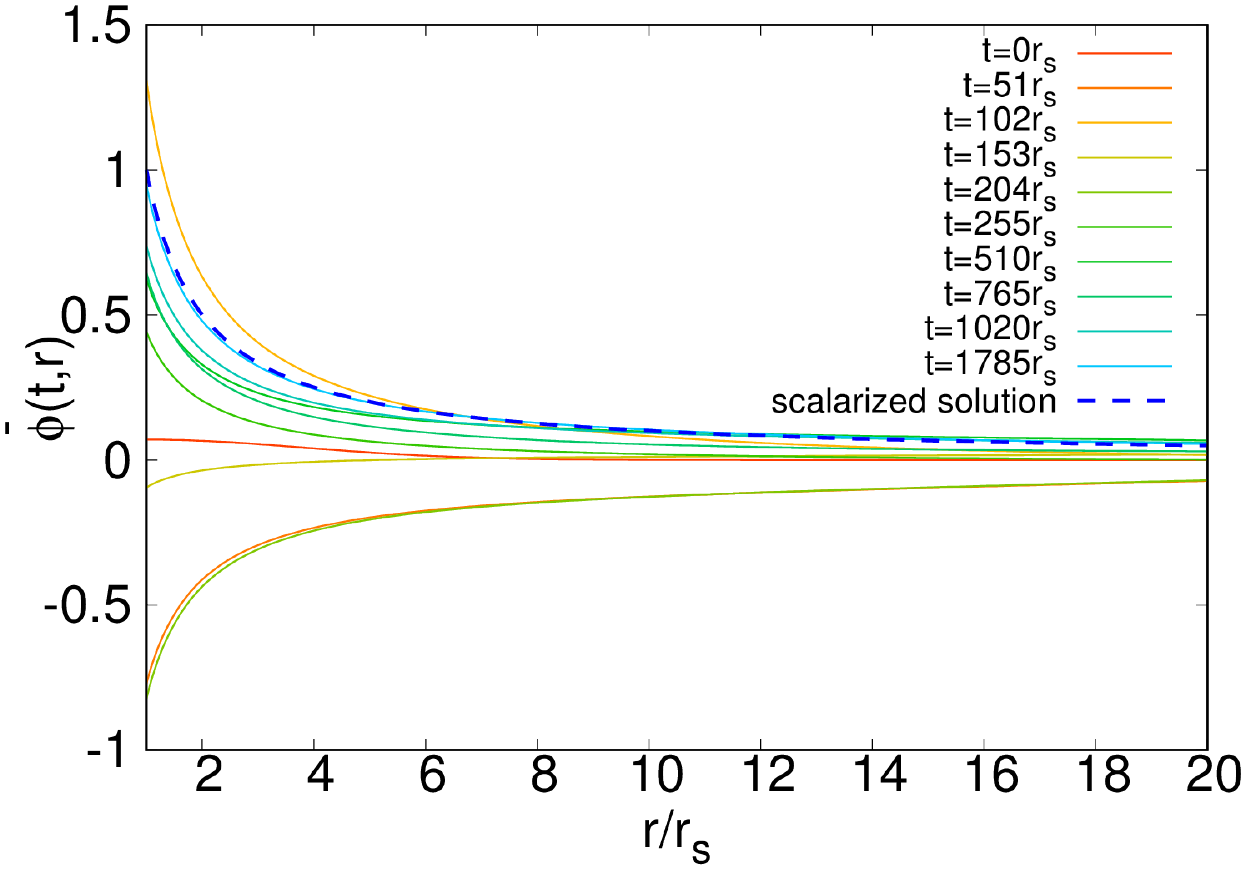}
\caption{Two illustrations of time evolution in the inverse quartic polynomial model~\eqref{model2} with a radiative boundary condition~\eqref{radiative BC1}-\eqref{radiative BC2}. The initial data are a momentarily static Gaussian initial data~\eqref{inidat} with $w=4r_{\rm s}$ and (top panels) $A\sqrt{\frac{ak^{2}}{2}}=0.01$, $\bar{Q}=\sqrt{2}$ or (bottom panels) $A\sqrt{\frac{ak^{2}}{2}}=0.07$, $\bar Q=0.1\sqrt{2}$. (Left panels) Contour plot of the time evolution for $r\bar{\phi}/r_{\rm s}$.
(Right panels) Radial profiles from snapshots of the time evolution at different times and of the final scalarized solution (blue dashed line) which has $\bar C_{0}\simeq 1.559$ (top right panel) or $\bar C_{0}\simeq 100$  (bottom right panel). 
}
\label{Graph_ms_A12_w4_r1_Q5_a01_b0005}
\centering
\end{figure}

Let us first report the evolutions for the inverse quartic polynomial model~\eqref{model2} and for the radiative boundary conditions~\eqref{radiative BC1}-\eqref{radiative BC2}. We have observed two qualitatively distinct types of evolution, both leading to a scalarized solution.  For large $\bar Q$ and a small initial perturbation, scalarization proceeds in a direct manner. That, is the tachyonic instability sets in, the scalar field grows, it briefly oscillates at each radial  position but quickly settles down to the scalarized solution. This is illustrated by the top panels in Fig.~\ref{Graph_ms_A12_w4_r1_Q5_a01_b0005}, for which 
the initial parameters are $(A,w)=(0.01\phi_{\rm min},4r_{s})$ and the charge is $\bar Q$ is $0.1\sqrt{2}$. 
The left panel shows that the scalarization starts in the vicinity of the sphere and propagates outwards. This is expected, as the strength of the tachyonic instability of the Coulomb solution decays with $r$. The right panel shows the radial profile, obtained from the evolution at different time slices, which can be observed to approach the one of a scalarized solution with $\bar C_{0}=1.559$, such that the relaxation time is longer for larger $r$, as expected. Moreover, the final value of  $\bar{\phi}$ at the conductor  is  $\bar{\phi}=1$ corresponding to an extremum of the effective potential, as expected from the radiative boundary condition (cf.~\eqref{extrema_qurt} and \eqref{redef}). Thus,  the final scalarized  solution interpolates (from the conductor to infinity)  between two extrema of the effective potential: $\bar{\phi}=1$ and $\bar{\phi}=0$.

The bottom panels in Fig.~\ref{Graph_ms_A12_w4_r1_Q5_a01_b0005} show a qualitatively different behaviour. After the tachyonic instability settles in, the scalar field initially evolves in  the ``wrong" direction; in this case it acquires negative values. Then, it oscillates around $\phi=0$, loosing energy and decreasing the oscillations amplitude. At some point, the scalar field stops crossing $\phi=0$, and starts to oscillate around its final non-trivial value (in this case a positive value), approaching the scalarized solution. This occurs for a smaller value  of  the charge,  $\bar Q=0.1\sqrt{2}$, and the scalarization process takes considerably longer. Hence, in the left bottom panel of Fig.~\ref{Graph_ms_A12_w4_r1_Q5_a01_b0005}, the slope is barely visible (but still present), indicating that, as before, scalarization proceeds faster in the immediate vicinity of the conductor, then spreading towards larger values of $r$. Again, the final value at the conductor  is  $\bar{\phi}=1$.

\begin{figure}[htbp]
\centering
\includegraphics[width=7.8cm]{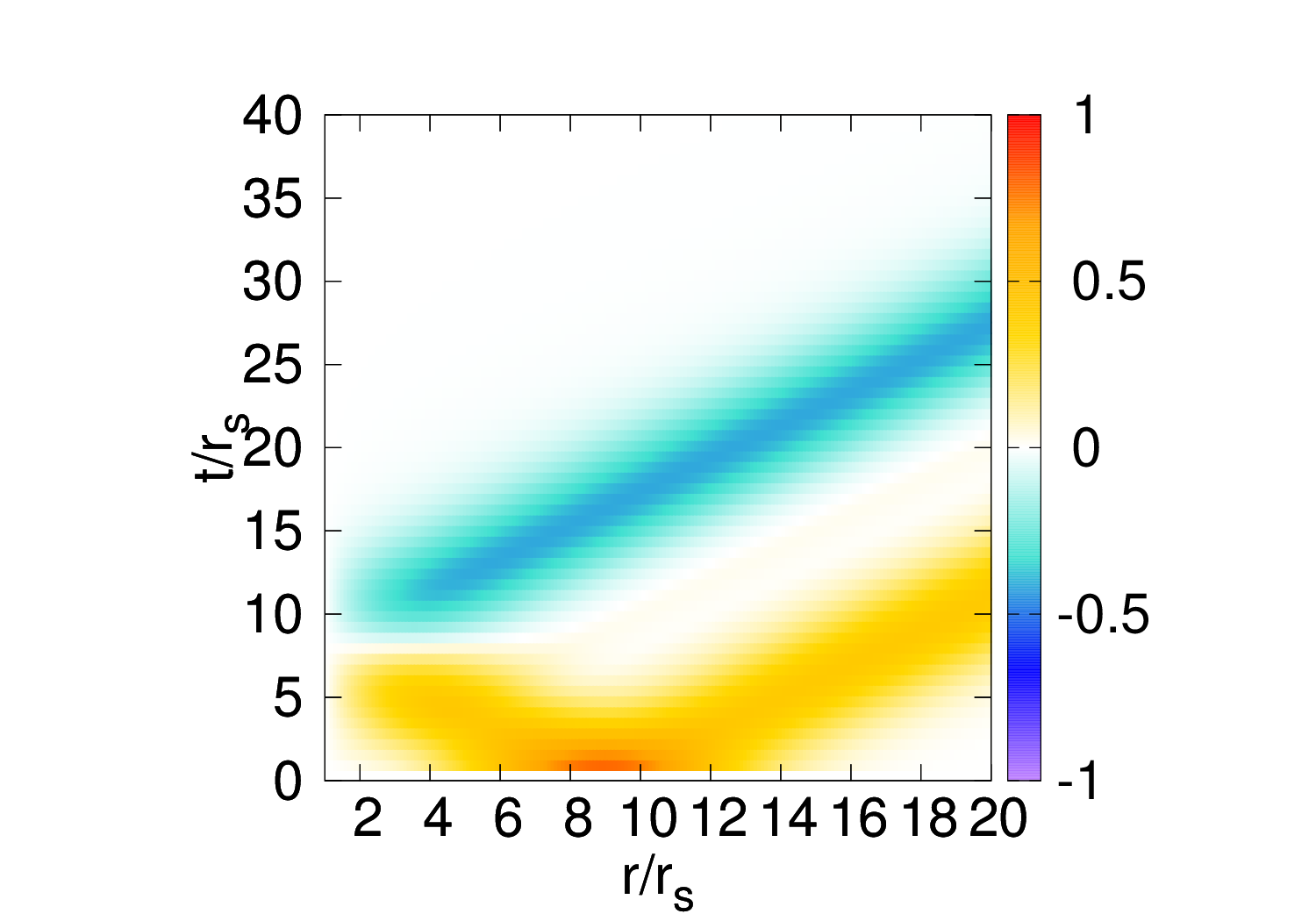} \\
\includegraphics[width=7.8cm]{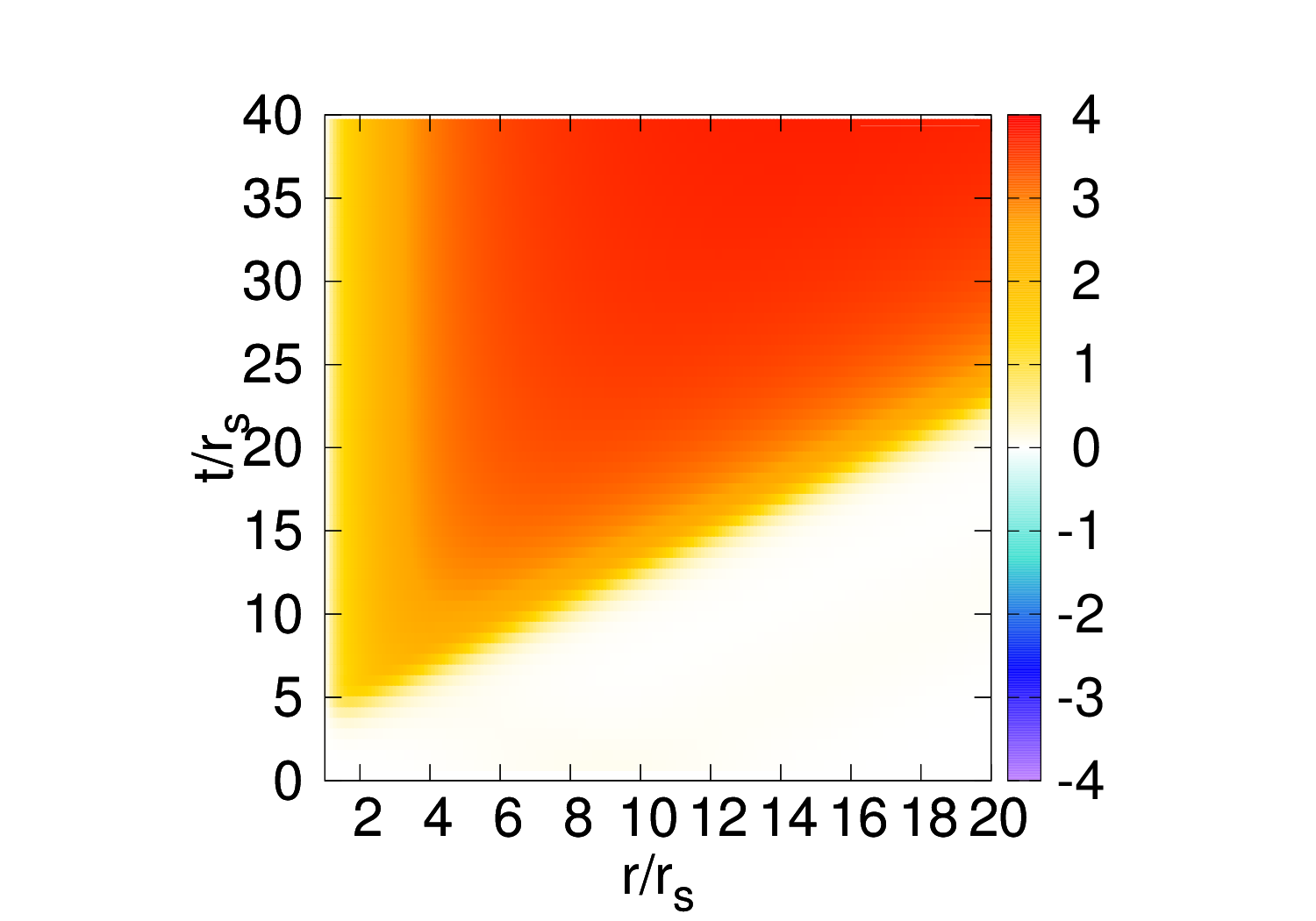}
\includegraphics[width=7cm]{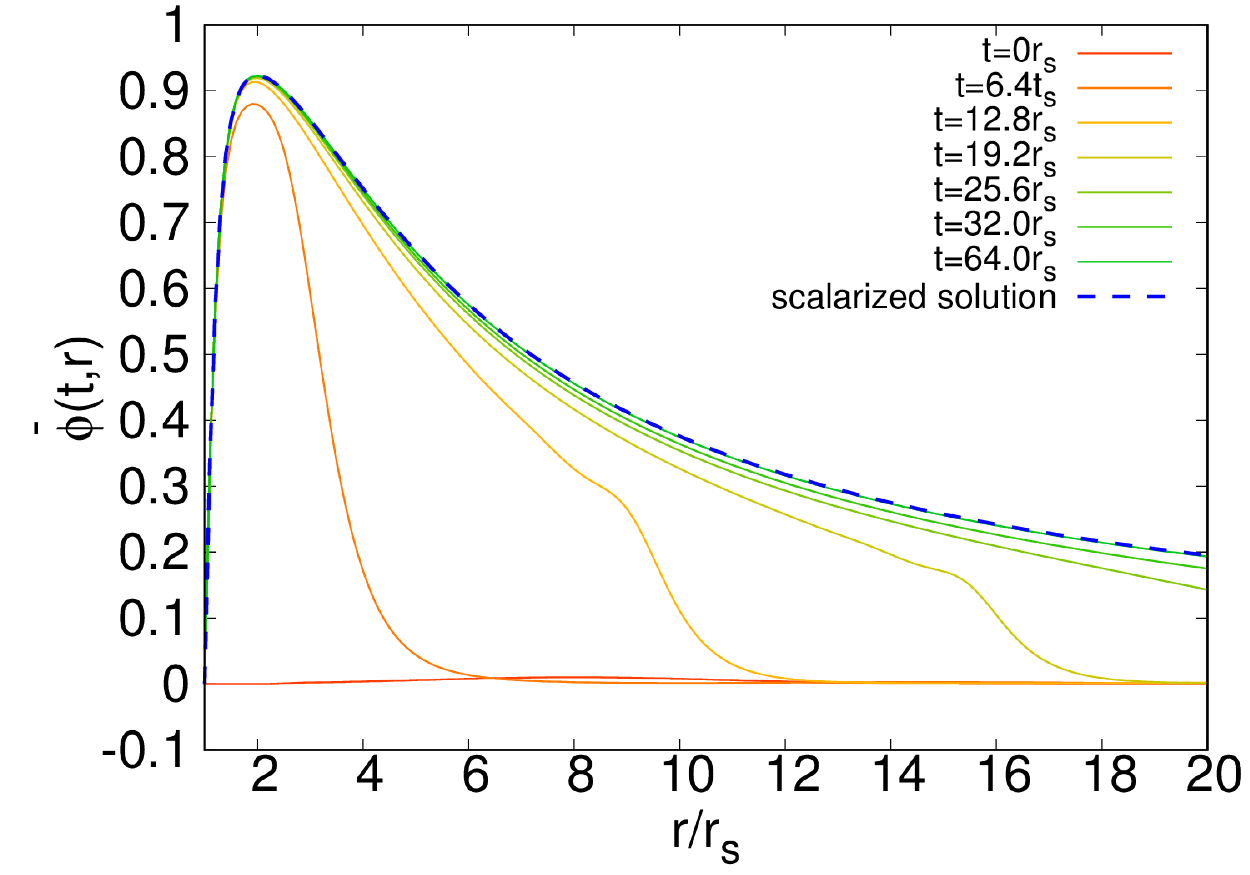}
\caption{Two illustrations of time evolution in the inverse quartic polynomial model~\eqref{model2} with 
the boundary condition (D)~\eqref{Dirichlet BC}. 
The initial data are a momentarily static Gaussian initial data~\eqref{inidat} with $w=4r_{\rm s}$, $r_{0}=8r_{\rm s}$, and (top panel)  $A\sqrt{\frac{ak^{2}}{2}}=0.1$,  $\bar{Q}=\sqrt{2}$ or (bottom panels)   $A\sqrt{\frac{ak^{2}}{2}}=0.01$, $\bar Q=4\sqrt{2}$. (Top and bottom left panels) Contour plot of the time evolution for $r\bar{\phi}/r_{\rm s}$.
(Bottom right panel) Radial profiles from snapshots of the time evolution at different times and of the final scalarized solution (blue dashed line) which has $\bar C_{0}\simeq 0.978$. In the simulation of the  top panel,   the final state is the Coulomb solution. 
}
\label{Graph_contour_time_evolution_ms_A01_w4_r8_QQ1_a2_b2_MD}
\centering
\end{figure}

Next, we consider evolution for the inverse quartic polynomial model~\eqref{model2} and for
the boundary condition (D)~\eqref{Dirichlet BC}. We shall also illustrate two qualitatively different types of behaviour. 
For sufficiently small charge, scalarization does not occur. 
This is illustrated by the simulation shown in the top panel of  Fig.~\ref{Graph_contour_time_evolution_ms_A01_w4_r8_QQ1_a2_b2_MD}.  
One observes that  the initial Gaussian perturbation produces some brief oscillations, but it essentially propagate outwards leaving behind an unscalarized Coulomb conductor.
The lack of scalarization for small charges agrees with the analysis in Section~\ref{sec2}.
For values of $\bar Q$ that can fit the stability window in Fig.~\ref{unstable}, 
we expect such a stable scalarized solution to form. 
This is illustrated by  the bottom panel of  Fig.~\ref{Graph_contour_time_evolution_ms_A01_w4_r8_QQ1_a2_b2_MD}. As in Fig.~\ref{Graph_ms_A12_w4_r1_Q5_a01_b0005} we can see scalarization to proceed first in the vicinity of the conductor  and then propagating outwards, this time in a direct manner, without oscillations. The endpoint can be identified as a scalarized solution with $\bar C_{0}\simeq 0.978$ (right panel). Observe that in this case the scalarized  solution has $\bar{\phi}=0$ at the conductor, and thus an extremum outside  the conducting  sphere, unlike the previous case, which is to be expected for 
the boundary conditions (D).

\begin{figure}[htbp]
\centering
\includegraphics[width=7.8cm]{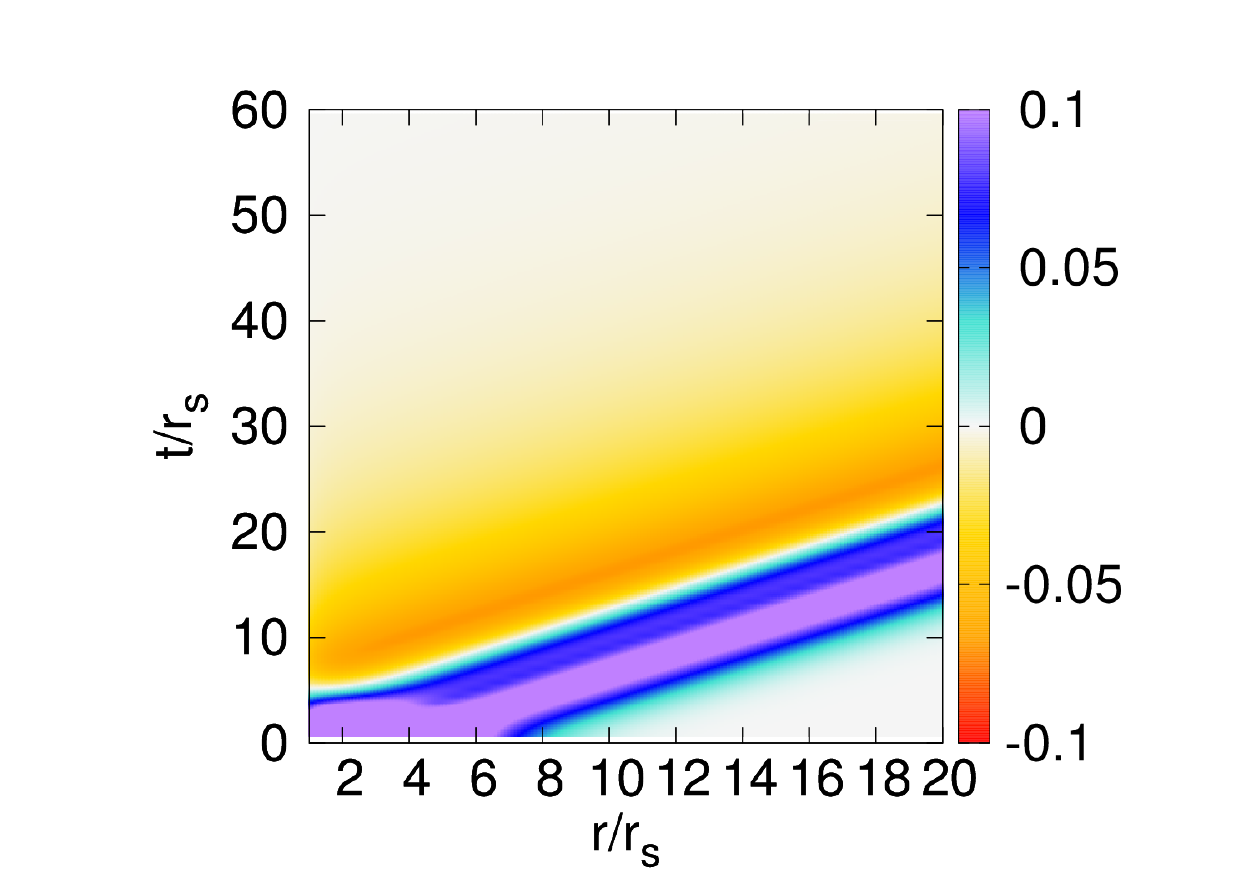}\\
\includegraphics[width=7.8cm]{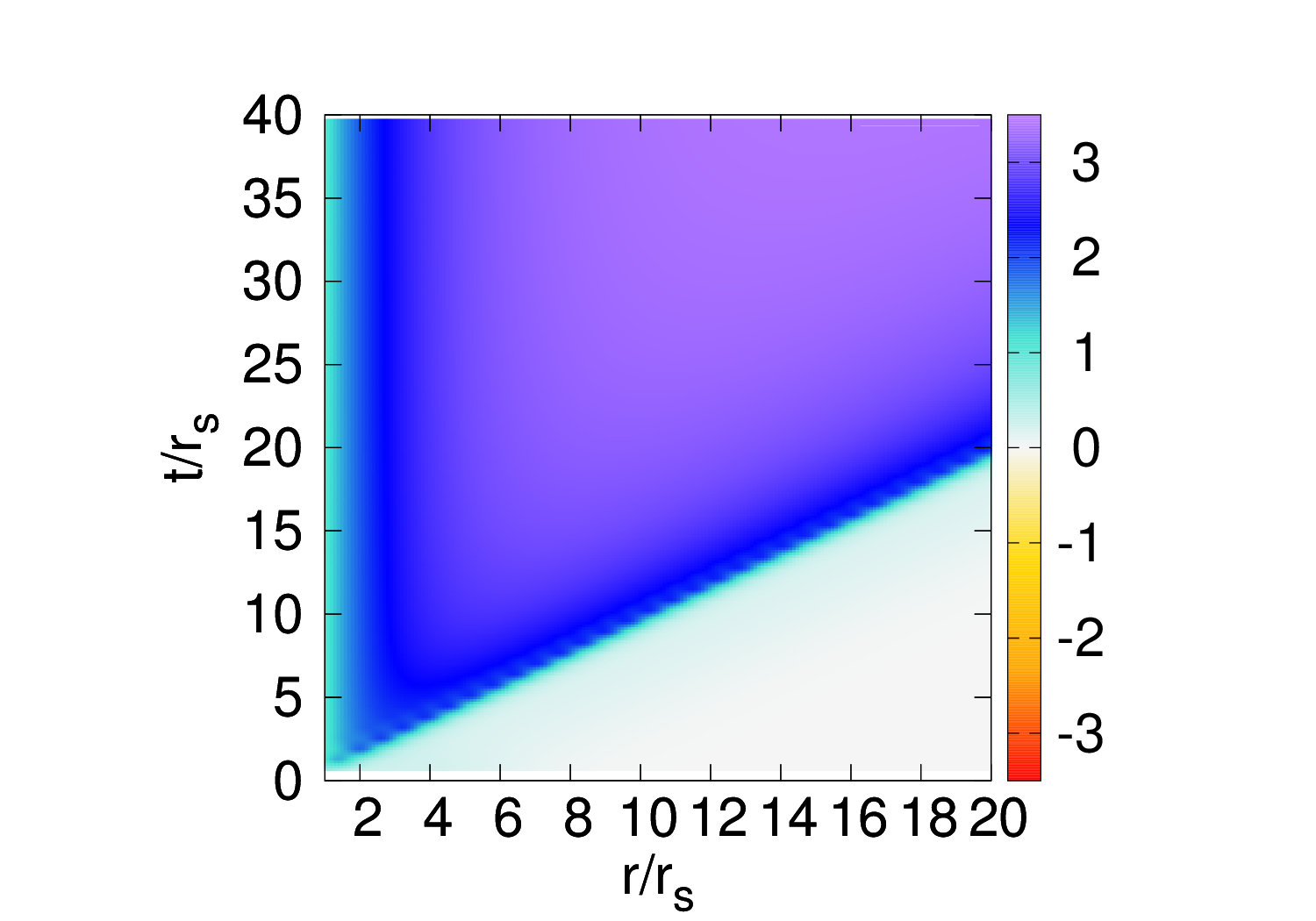}
\includegraphics[width=7cm]{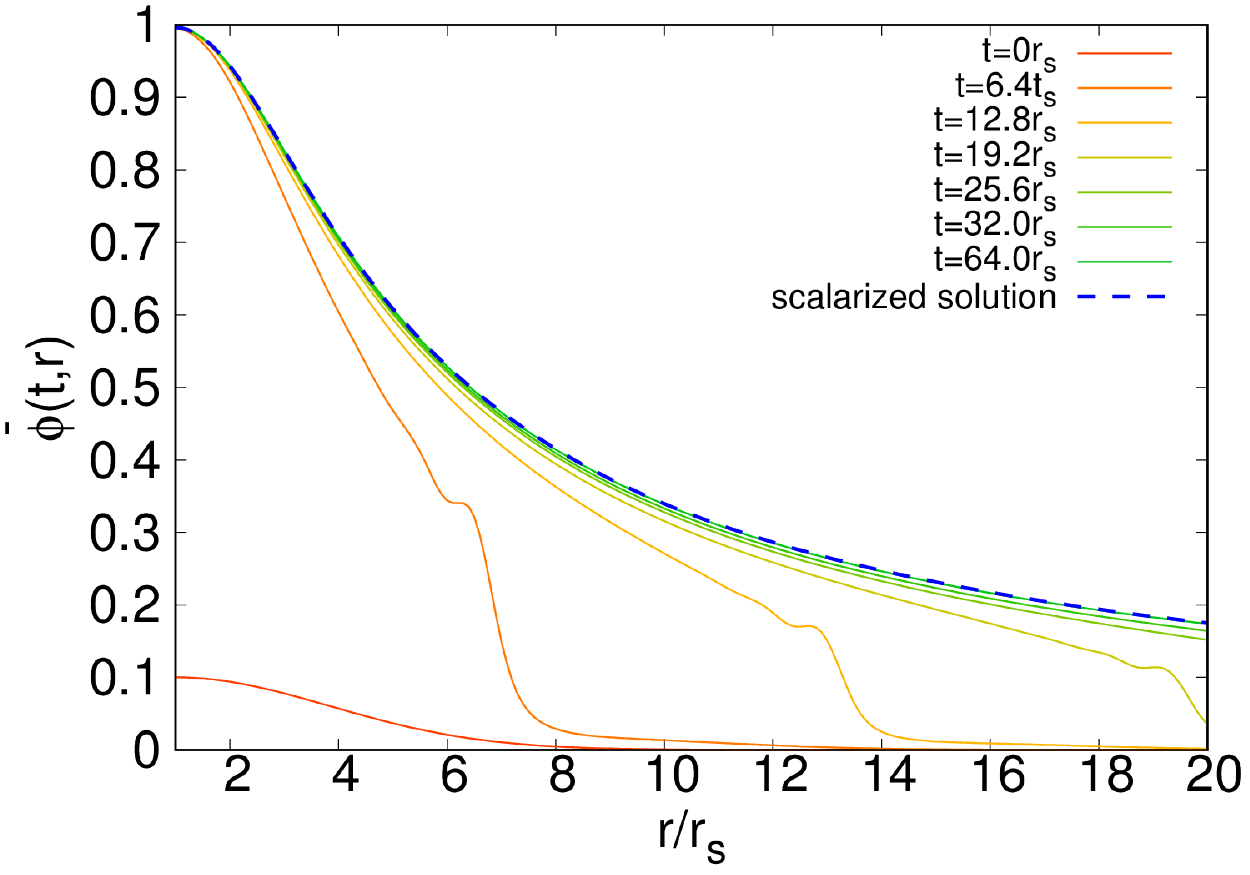}
\caption{
Two illustrations of time evolution in the inverse quartic polynomial model~\eqref{model2} with
the boundary condition (N)~\eqref{Neumann BC}.
The initial data are a momentarily static Gaussian initial data~\eqref{inidat} with $r_{0}=r_{\rm s}$ and (top panel) $A\sqrt{\frac{ak^{2}}{2}}=0.1$, $w=4r_{\rm s}$, $\bar Q=1$ or (bottom panels) $A\sqrt{\frac{ak^{2}}{2}}=0.1$, $w=4r_{\rm s}$, $\bar Q=5$. (Top and bottom left panels) Contour plot of the time evolution for $r\bar{\phi}/r_{\rm s}$.
(Bottom right panel) Radial profiles from snapshots of the time evolution at different times and of the final scalarized solution (blue dashed line) which has $\bar{C}_{0}\simeq 0.99995$. In the simulation of the  top panel,   the final state is the Coulomb solution.  
}
\label{Graph_contour_time_evolution_ms_A01_w4_r1_QQ1_a2_b2_MN}
\centering
\end{figure}

Finally, we consider evolutions for the inverse quartic polynomial model~\eqref{model2} and for the boundary condition (N)~\eqref{Neumann BC}. 
The situation is analogous to that of the
boundary condition (D) and 
we shall illustrate the same two qualitatively different behaviours. For a sufficiently small charge, scalarization does not occur. This is illustrated by the simulation shown in the top panel of  Fig.~\ref{Graph_contour_time_evolution_ms_A01_w4_r1_QQ1_a2_b2_MN}.  As in the previous case,  the initial Gaussian perturbation produces some brief oscillations, but then dissipates away. This is in agreement with the analysis in Section~\ref{sec2}. But for values of $\bar Q$ that can fit the stability window in Fig.~\ref{unstable}, a scalarized solution forms. This is illustrated by  the bottom panel of  Fig.~\ref{Graph_contour_time_evolution_ms_A01_w4_r1_QQ1_a2_b2_MN}. 
As in the 
(D) case, scalarization proceeds in a rather monotonic fashion, without oscillations. 
Unlike the (D) 
case, however, the scalarized  solution has no extremum outside  the conducting  sphere and has $\bar{\phi}=1$ on the conductor, again, interpolating between the two extrema of the potential at the conductor ($\bar{\phi}=1$) and infinity ($\bar{\phi}=0$).

\subsubsection{Inverse cosine model}
\begin{figure}[htbp]
\centering
\includegraphics[width=7.8cm]{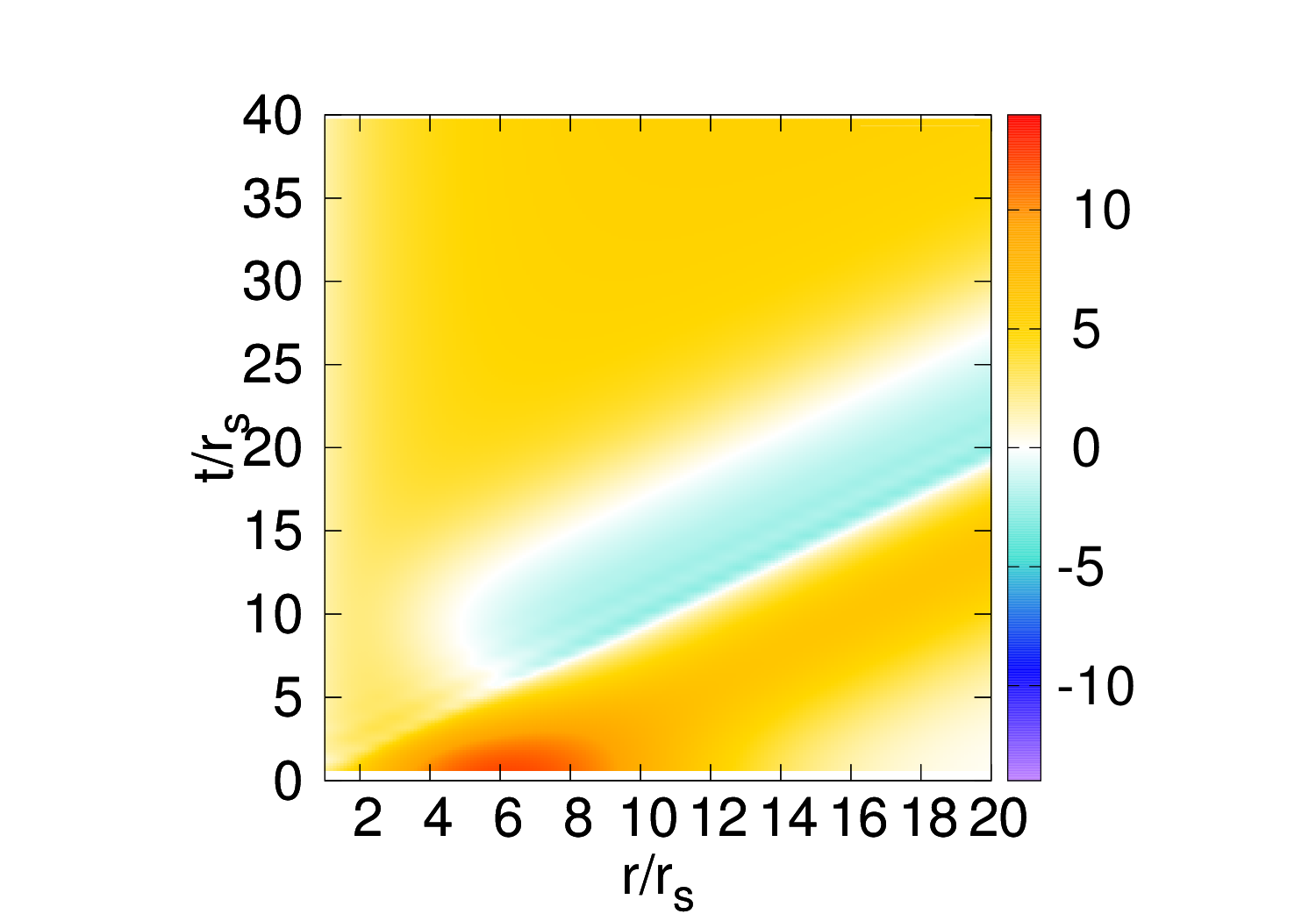}
\includegraphics[width=7cm]{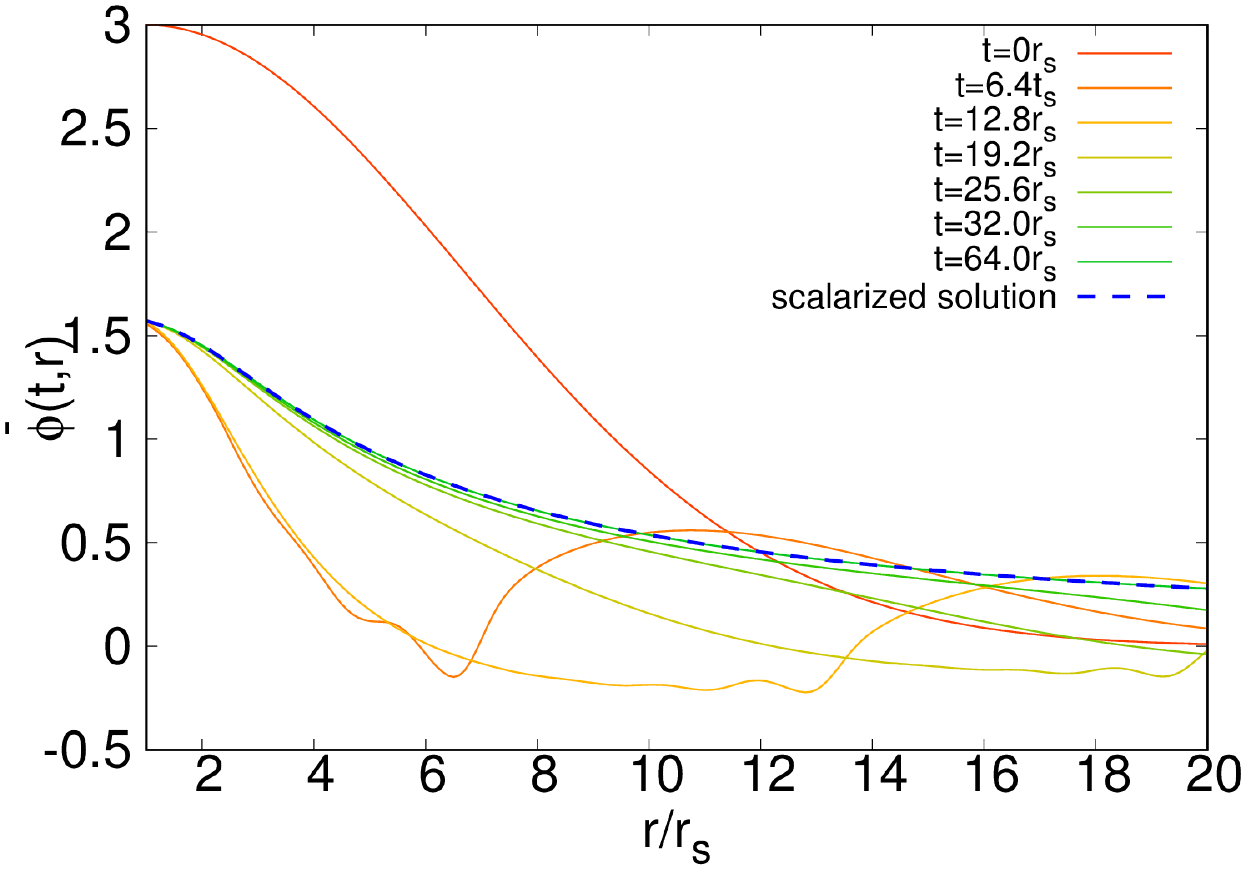}\\
\includegraphics[width=7.8cm]{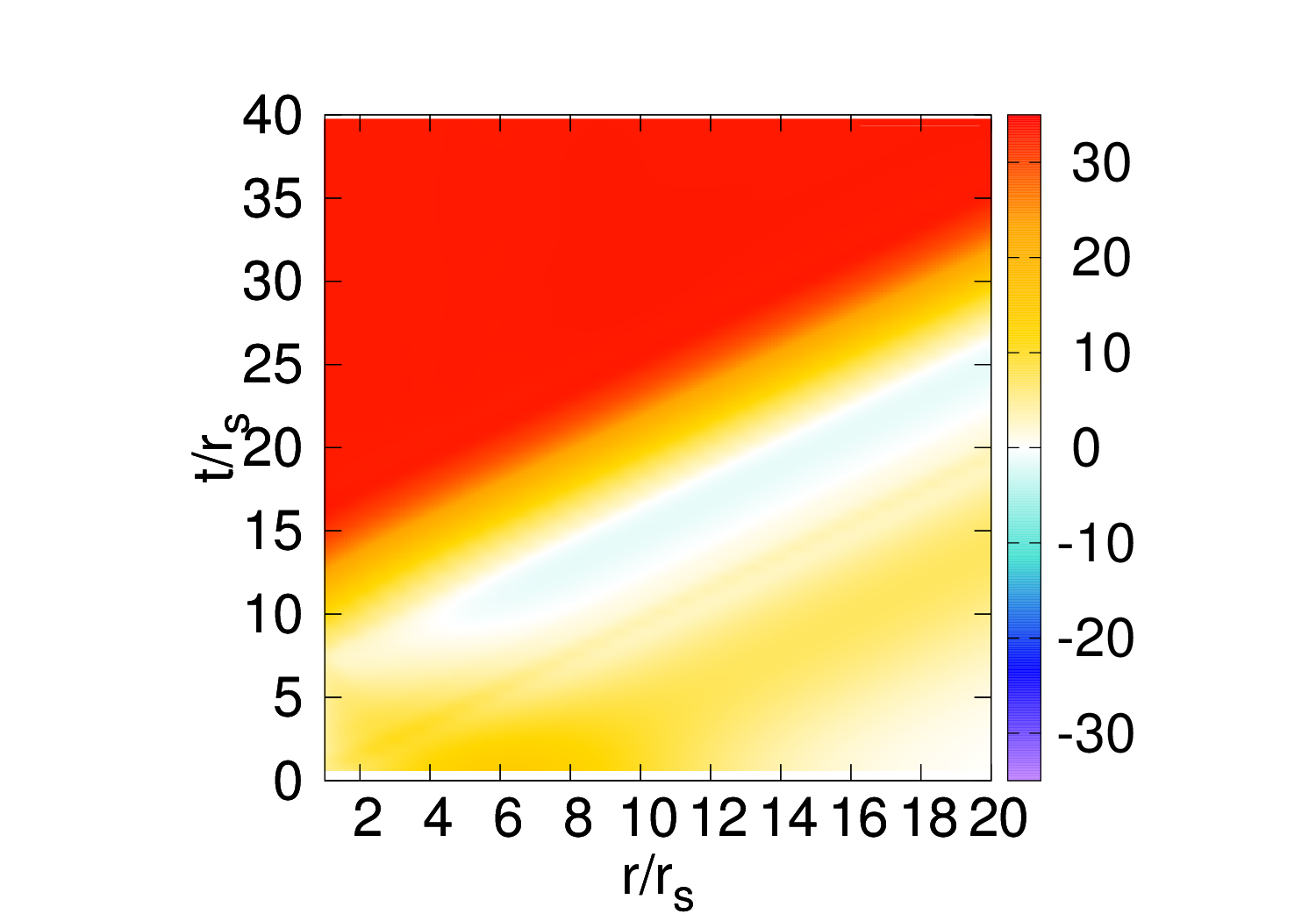}
\includegraphics[width=7cm]{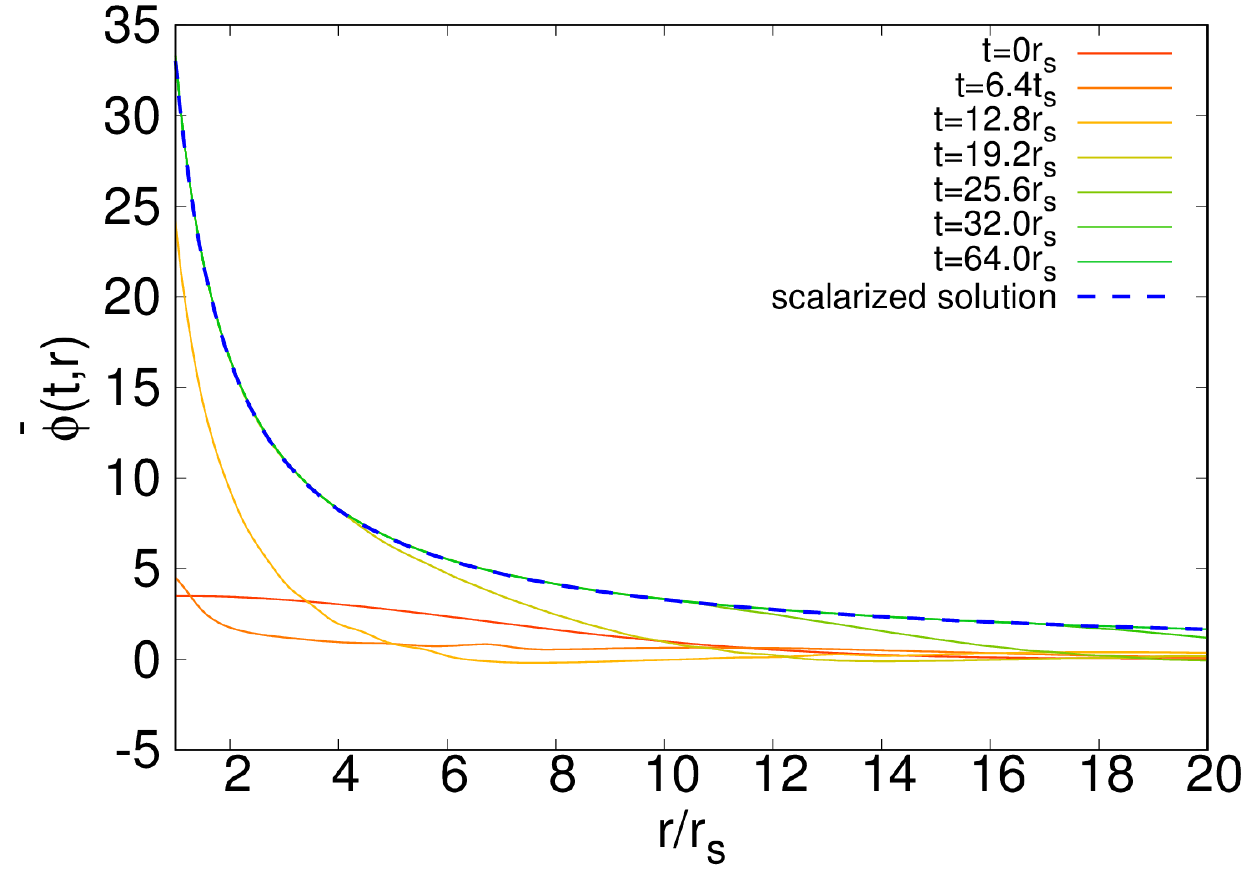}
\caption{
Two illustrations of time evolution in the inverse cosine model~\eqref{model3} with a radiative boundary condition~\eqref{radiative BC1}-\eqref{radiative BC2}. The initial data are a momentarily static Gaussian initial data~\eqref{inidat} with $r_{0}=r_{\rm s}$, $w=8.0r_{s}$, $\bar{Q}=4\sqrt{2}$ and (top panels) $A\sqrt{\frac{a}{2}}=3.0$, or (bottom panels) $A\sqrt{\frac{a}{2}}=3.5$. (Left panels) Contour plot of the time evolution for $r\bar{\phi}/r_{\rm s}$.
(Right panels) Radial profiles from snapshots of the time evolution at different times and of the final scalarized solution (blue dashed line) which has $C_{1}\simeq 2.0004$ (top right panel) or $C_{1}\simeq 69.0$  (bottom right panel). 
}
\label{fig9}
\centering
\end{figure}

\begin{figure}[htbp]
\centering
\includegraphics[width=7.8cm]{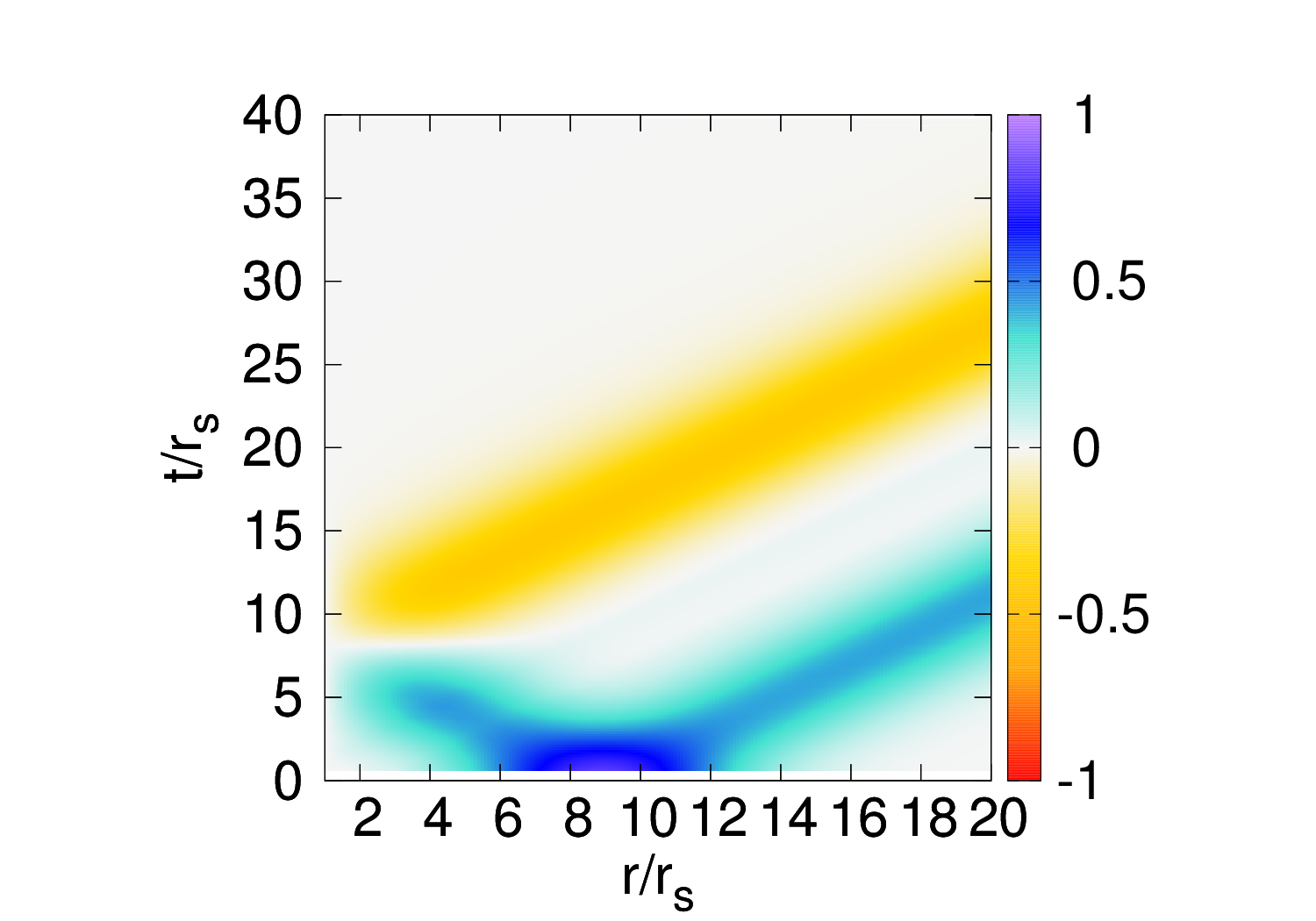}\\
\includegraphics[width=7.8cm]{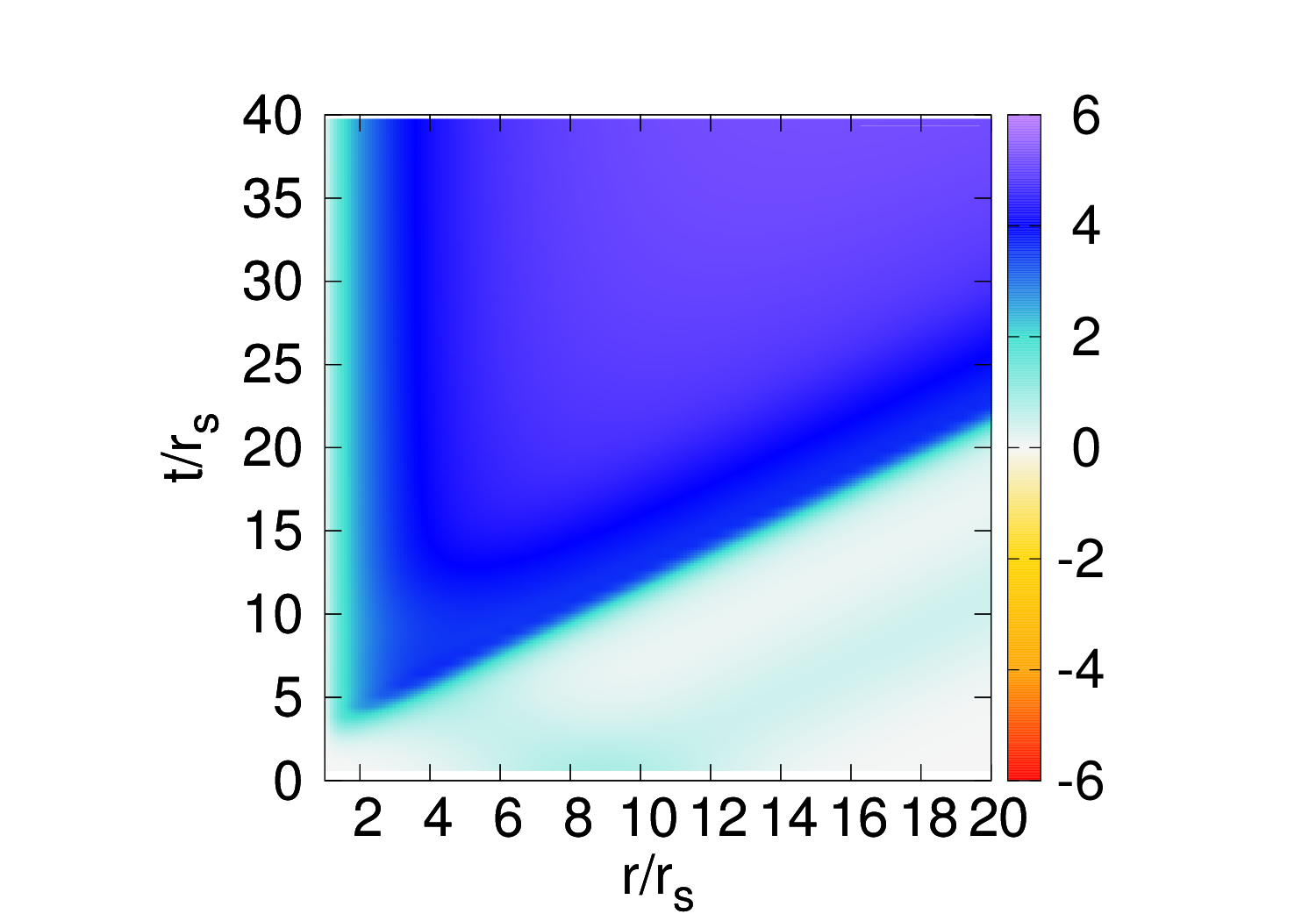}
\includegraphics[width=7cm]{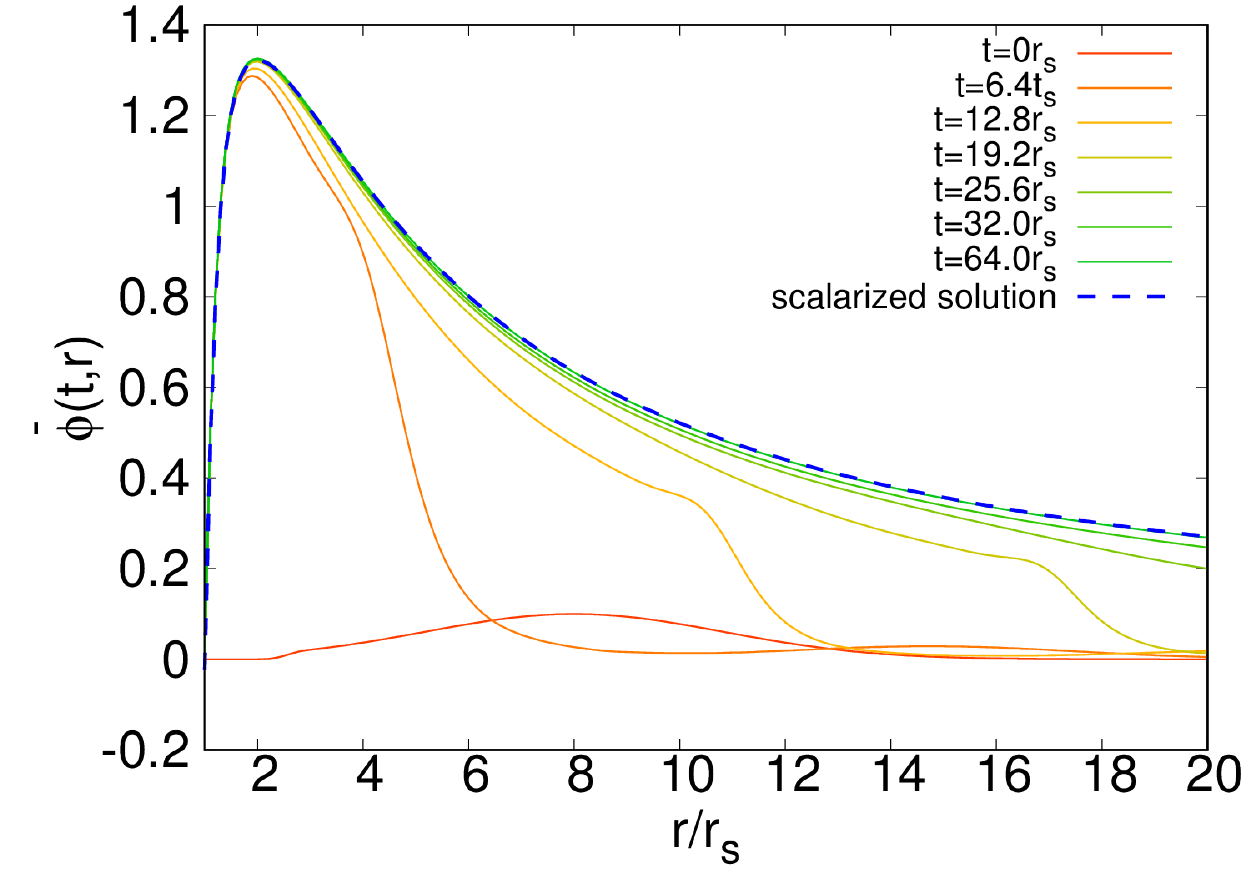}
\caption{
Two illustrations of time evolution in the inverse cosine model~\eqref{model3} with 
the boundary condition (D)~\eqref{Dirichlet BC}. The initial data are a momentarily static Gaussian initial data~\eqref{inidat} with $r_{0}=8r_{\rm s}$ and (top panel) $A\sqrt{\frac{a}{2}}=0.1$, $w=4r_{\rm s}$, $\bar Q=\sqrt{2}$ or (bottom panels) $A\sqrt{\frac{a}{2}}=0.1$, $w=4r_{\rm s}$, $\bar Q=4\sqrt{2}$. (Top and left bottom panels) Contour plot of the time evolution for $r\bar{\phi}/r_{\rm s}$.
(Right bottom panel) Radial profiles from snapshots of the time evolution at different times and of the final scalarized solution (blue dashed line) which has $C_{1}\simeq 1.88$. 
In the simulation of the  top panel,  the final state is the Coulomb solution.
}
\label{fig10}
\centering
\end{figure}

\begin{figure}[htbp]
\centering
\includegraphics[width=7.8cm]{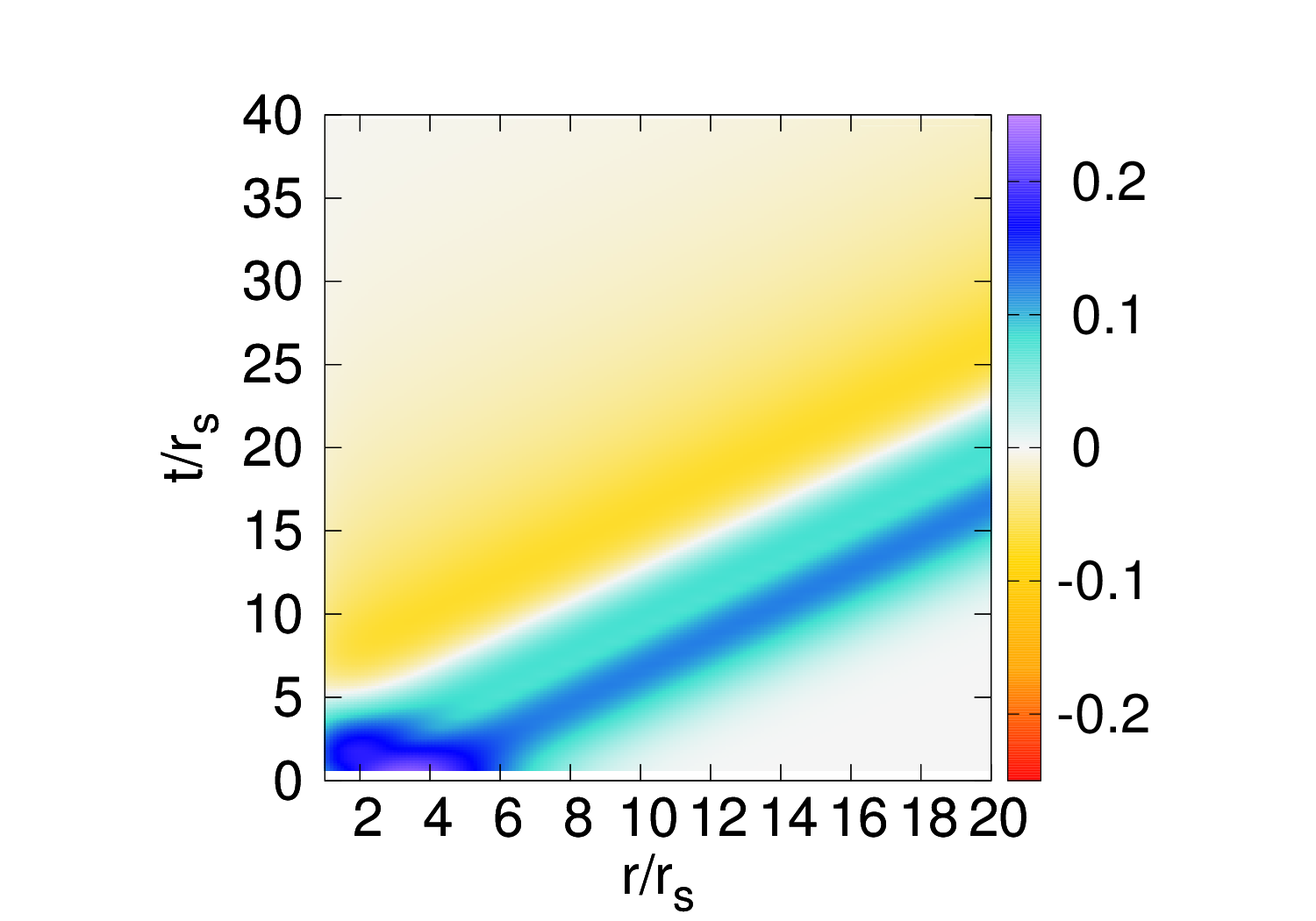}\\
\includegraphics[width=7.8cm]{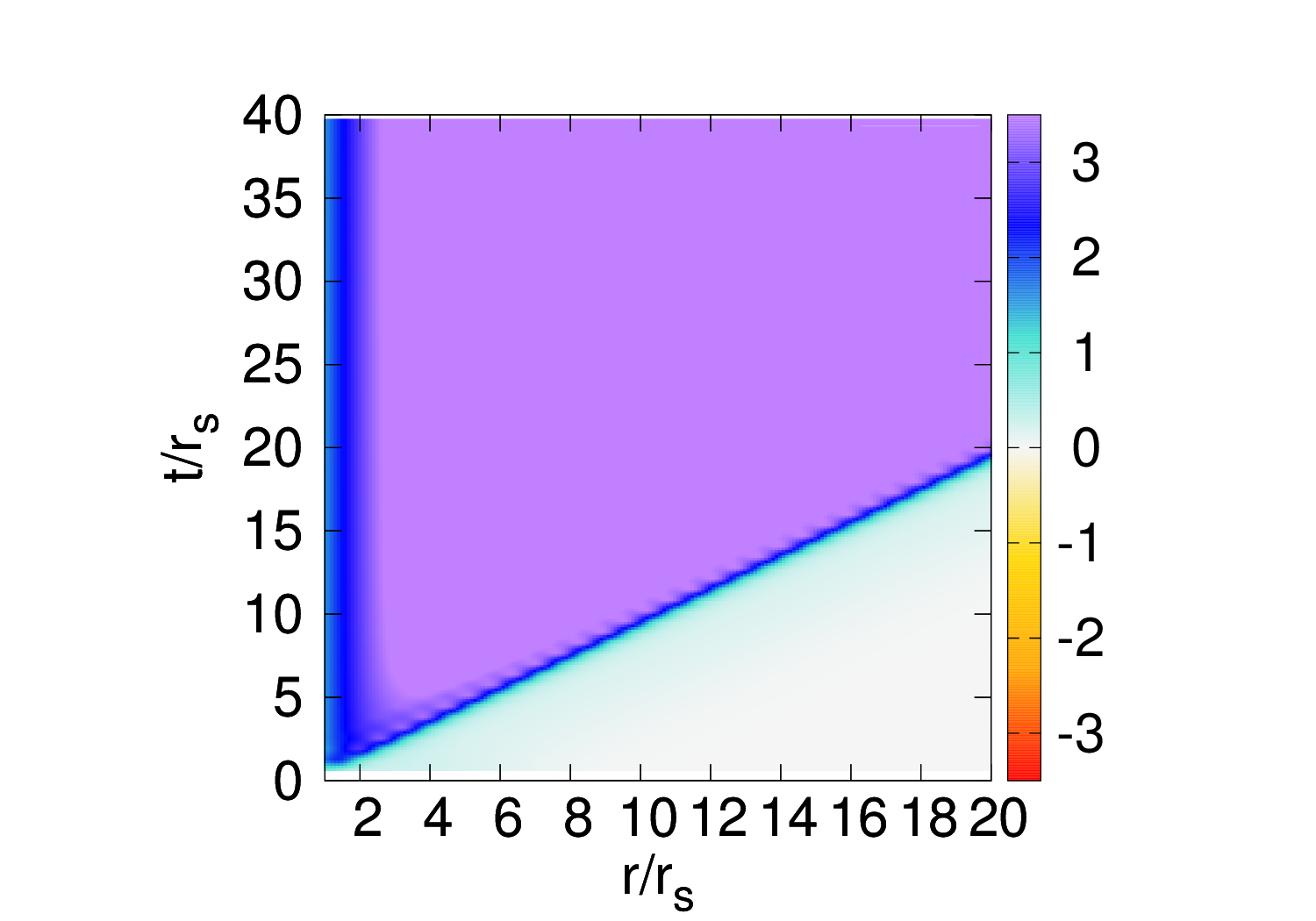}
\includegraphics[width=7cm]{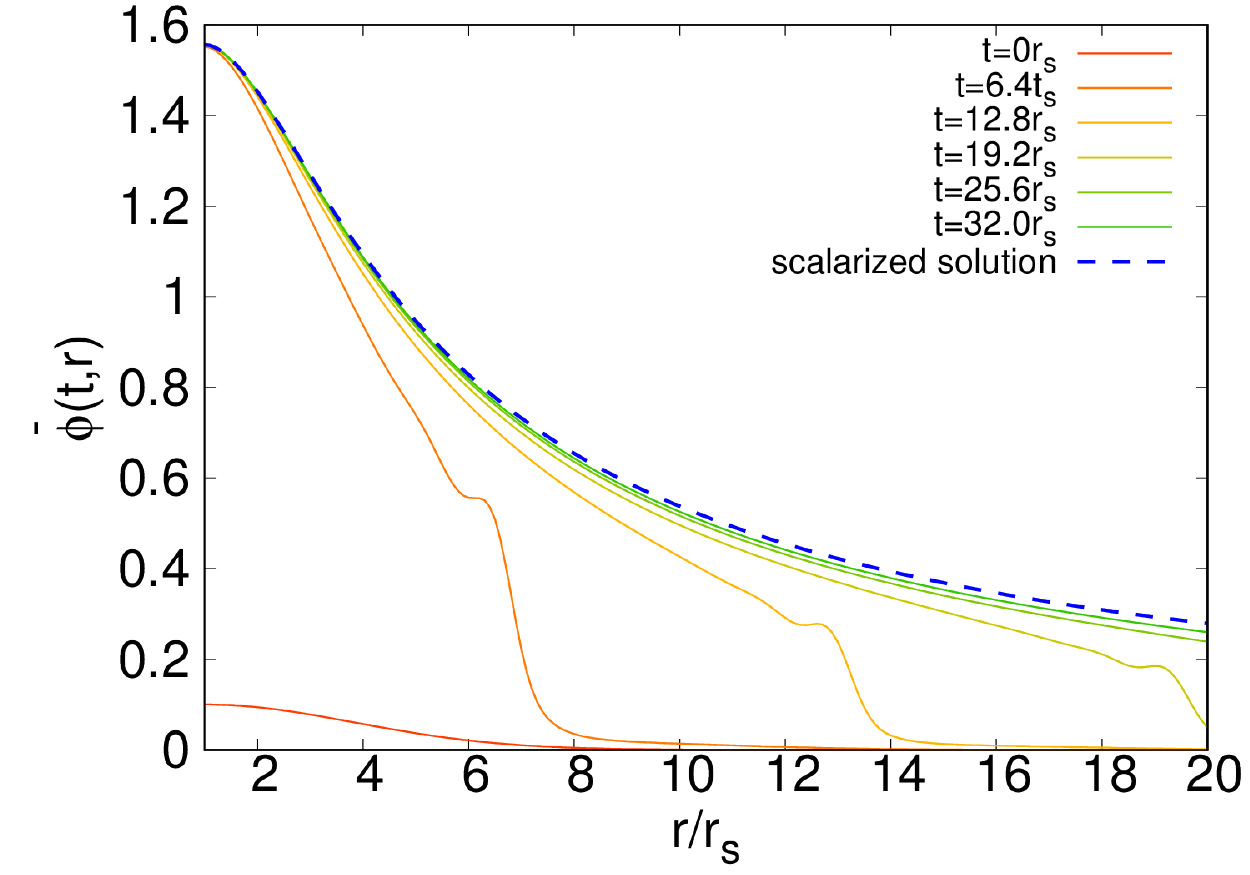}
\caption{
Two illustrations of time evolution in the inverse cosine model~\eqref{model3} with
the boundary condition (N)~\eqref{Neumann BC}. The initial data are a momentarily static Gaussian initial data~\eqref{inidat} with $r_{0}=r_{\rm s}$, $A\sqrt{\frac{a}{2}}=0.1$, $w=4r_{\rm s}$ and (top panel) $\bar Q=\sqrt{2}$ or (bottom panels) $\bar Q=4\sqrt{2}$. (Top and left bottom panels) Contour plot of the time evolution for $r\bar{\phi}/r_{\rm s}$.
(Right bottom panel) Radial profiles from snapshots of the time evolution at different times and of the final scalarized solution (blue dashed line) which has $C_{1}\simeq 1.9996$. In the simulation of the  top panel,   the final state is the Coulomb solution.
}
\label{fig11}
\centering
\end{figure}

The evolution for the inverse cosine model~\eqref{model3} shows qualitatively similar features to the ones of the inverse quartic polynomial model~\eqref{model2} discussed in the previous subsection. These are illustrated for the radiative boundary conditions~\eqref{radiative BC1}-\eqref{radiative BC2} in Fig.~\ref{fig9}, for 
the boundary condition (D) in Fig.~\ref{fig10} and for 
the boundary condition (N) in  Fig.~\ref{fig11}.
However, as is shown in Fig.~\ref{unstable}, there are many stable scalarized solutions for the inverse cosine model with radiative boundary condition.
This fact suggests that the final state may depend on the initial data.
Fig.~\ref{fig9} shows the two simulations with same $\bar{Q}$ and different initial data, and the final scalarized solution is different.
The simulations with the
boundary conditions (D) and (N) in  Figs.~\ref{fig10}-\ref{fig11}, each considers one case where scalarization occurs (bottom panel) and one where it does not (top panel).

\section{Conclusions}
\label{sec6}
In this paper we have investigated spontaneous scalarization of a charged, conducting sphere  in Maxwell-scalar~(Ms) models. These models provide  a simple, yet sufficiently rich, arena to study the spontaneous scalarization phenomenon, which has recently received considerable attention in the context of strong gravity. Indeed, as we have explained, Ms models can parallel features observed in the gravitational context, such as the coupling dependence on the existence (or not) of  stable scalarized solutions. We believe that, in this respect, Ms models are insightful on the  behaviour of the strong gravity models. On  the other hand, Ms models for scalarization of a conductor have features absent in strong gravity, such as the freedom to choose boundary conditions on  the conductor, which  in the  gravitational case is fixed by requiring regularity  at the event horizon. We have explored how the choice of boundary conditions impacts on the  physical properties of the system, showing that such impact can be of considerable relevance. For instance, in the case of radiative boundary condition, the Coulomb solution is always unstable against scalarization, as shown in Section~\ref{sec2}. 
This is  not the case for Dirichlet or Neumann  boundary conditions, for which there are regions of stability and regions of instability.

The simplicity of the Ms model allows also to perform numerical evolution of the scalarization phenomenon which is more challenging in the strong gravity case, although  they have been performed for charged black holes in, e.g.,~\cite{Herdeiro:2018wub,Fernandes:2019rez}. The simulation herein helps us to understand how scalarization proceeds, when it occurs. We can distinguish two main behaviours:
(i) direct scalarization; 
the scalar field directly grows to form the scalarized solution, 
from the Coulomb solution; 
(ii) oscillating scalarization; 
the scalar field oscillates between the two scalarized solutions before settling into one of the solutions. If the charge is sufficiently small, on the other hand, the Coulomb solution is stable against scalarization (except for radiative boundary conditions).

For black hole scalarization in scalar-gravity systems, the boundary conditions in the limit of the event horizon are determined by regularity,
and the derivative of the scalar field in the horizon limit 
is automatically fixed once the horizon field value is specified,
and thus the flexibility seen in this paper has no parallelism,
However, scalar-gravity systems allow the scalarization of sufficiently compact objects, 
albeit non-black holes, such as neutron stars.  
In this case, there will be a similar freedom in the choice of boundary conditions; for instance, in the case of neutron stars the derivative of the scalar field at the center of the star 
also depends on the energy density and pressure at the center of the star
obeying the given equation of state.
Therefore, one of the lessons from this work is that neutron star scalarization may depend on the  choice of the boundary condition on the surface of such a compact object.

Adding the mass term $2m^2 \phi^2$ or self-interaction potential $V(\phi)$
in the action \eqref{action}
may significantly modify our results.
As in the case of spontaneous scalarization
of neutron stars \cite{Ramazanoglu:2016kul,Yazadjiev:2016pcb,AltahaMotahar:2019ekm}
and black holes \cite{Macedo:2019sem},
adding the mass term to the  action \eqref{action} may change the asymptotic behaviour of the scalar field,
while
it may be possible to stabilize scalarized solutions even in the linear model 
by including appropriate self-interactions.
The investigation of these generalized models is, however, beyond the scope of this paper.

Finally, as a direction for further research, the simplicity of the Ms model may allow us to consider spontaneous scalarization  beyond spherical symmetry, analysing, in particular, the role of dissipation played by the electromagnetic radiation.

\bigskip

{\bf Acknowledgements.}
This work is supported 
by the Center for Research and Development in Mathematics and Applications (CIDMA) 
through the Portuguese Foundation for Science and Technology (FCT - Fundac\~ao para a Ci\^encia e a Tecnologia), references UIDB/04106/2020, UIDP/04106/2020, and by national funds (OE), through FCT, I.P., in the scope of the framework contract foreseen in the numbers 4, 5 and 6 of the article 23, of the Decree-Law 57/2016, of August 29, changed by Law 57/2017, of July 19.
We wish to thank for the financial support from the CENTRA through the Project~No.~UIDB/00099/2020 (FCT).
We also acknowledge support from the Projects No. PTDC/FIS- OUT/28407/2017, No. CERN/FIS-PAR/0027/2019 and No. PTDC/FIS-AST/3041/2020. 
This work has further been supported by the European Union’s Horizon 2020 research and innovation (RISE) programme H2020-MSCA-RISE-2017 Grant No. FunFiCO-777740.  The authors would like to acknowledge networking support by the COST Action CA16104.
This work is also supported by the European Union’s H2020 ERC Consolidator Grant ”Matter and strong-field gravity: New frontiers in Einstein’s theory” grant agreement no. MaGRaTh- 646597, under the H2020-MSCA-RISE-2015 Grant No. StronGrHEP-690904, and under the European Union's H2020 ERC, Starting Grant agreement no.~DarkGRA--757480 (T.I.).
Computations were performed on the “Baltasar Sete-Sois” cluster at IST and XC40 at YITP in Kyoto University.
T. N. acknowledges financial support from rigaku wakate kaigai haken program in Nagoya university.

\bigskip

\appendix
\section{The radial stability analysis in Ms models}
\label{appendix1}

To assess the linear stability of the solutions in Ms models
given by Eqs.~\eqref{quadratic} and~\eqref{quadratic2}, 
we consider small radial perturbations given by Eq. \eqref{radial_perturbations}.
We then expand the field equations up to first order in $\epsilon$,
and obtain the perturbed scalar field equation given by 
\be
\label{scalar_pert_eq}
-\ddot{\phi}_1
+\phi_1''
+\frac{2}{r}\phi_1'
+\frac{Q^2f_{\phi\phi}(\phi_0)}
         {2r^4 f(\phi_0)^2}
\phi_1
-\frac{Qf_\phi(\phi_0) }{r^2 f(\phi_0)}
\left(
\dot{a}_r-a_t'
\right)
=0 \ .
\ee
Integrating the $t$-component of the 
the perturbed vector field equation
provides the relation
\be
\dot{a}_r
=a_{t}'
+\frac{Q f_\phi (\phi_0)}
          {r^2 f(\phi_0)^2} 
\phi_1 \ ,
\ee
where we set the integration constant  to be zero; 
using the last relation in the perturbed scalar field equation \eqref{scalar_pert_eq} yields
\begin{eqnarray}
-\ddot{\phi}_1
+\phi_1''
+\frac{2}{r}\phi_1'
+\frac{Q^2}
         {2r^4 f(\phi_0)^3}
\left[
f (\phi_0)f_{\phi\phi}(\phi_0)
-2f_\phi(\phi_0)^2
\right]
\phi_1
=0 \ .
\end{eqnarray}
Introducing $\Phi:=r\phi_1$,
one can eliminate the first derivative terms
\be
\Phi''
+\frac{Q^2}
         {2r^4 f(\phi_0)^3}
\left[
f (\phi_0)f_{\phi\phi}(\phi_0)
-2f_\phi(\phi_0)^2
\right]
\Phi
=
\ddot{\Phi}\ .
\ee
Fourier decomposing the field $\Phi$, 
$\Phi=\int d\omega \Phi_\omega e^{-i\omega t}$,
each mode obeys a time independent Schr\"odinger-type equation
\be
-\Phi_\omega''
+ V_{\rm eff}(r)
\Phi_\omega
=
\omega^2 \Phi_\omega \ ,
\ee
where the effective potential $V_{\rm eff}(r)$ is given by Eq. \eqref{eff_potential}.

\section{On the unstable modes}
\label{appendix2}

Since the linear stability analysis depends on the chosen boundary conditions, in this appendix
we revisit the standard argument of the analysis for our setup.
We start from Eq. \eqref{dimless_Sch}
with a time independent Schr\"odinger-type form
\be
-\Phi_\omega''
+ V_{\rm eff}(\bar{r})
\Phi_\omega
=
\bar{\omega}^2 \Phi_\omega \ ,
\ee
where `prime' here denotes the derivative with respect to ${\bar r}$.
Multiplying $\Phi_{\omega}^{\ast}$ and integrating from the surface of the conductor to infinity, we have
\be
\bar{\omega}^{2}
\int_{1}^{\infty}
d{\bar r}
\left|
\Phi_{\omega}
\right|^{2}
&=&
-
\Phi_{\omega}^{\ast}\Phi_{\omega}'
\Big|_{r\to \infty}
+
\Phi_{\omega}^{\ast}\Phi_{\omega}'
\Big|_{\bar{r}\to 1}
+\int^{\infty}_{1}
d{\bar r}
\left[
\left|
\Phi_{\omega}'
\right|^{2}
+
V_{\rm eff}(\bar{r})\left|
\Phi_{\omega}
\right|^{2}
\right]\,
\nonumber\\
&=&
-i\bar{\omega}
\left|\Phi_{\omega} (\infty)\right|^2
+
\Phi_{\omega}^{\ast}\Phi_{\omega}'
\Big|_{\bar{r}\to 1}
+\int^{\infty}_{1}d\bar{r}
\left[
\left|
\Phi_{\omega}'
\right|^{2}
+
V_{\rm eff}(\bar{r})\left|
\Phi_{\omega}
\right|^{2}
\right]\,,
\label{eq:S deformation}
\ee
where 
we have imposed the outgoing boundary condition at spatial infinity:
\be
\Phi_{\omega}'(\infty)
\to i\bar{\omega} \Phi_\omega (\infty) \,.
\ee
On the other hand, 
as in Eqs. \eqref{eq:ingoing condition} and \eqref{eq:dirichlet condition},
we consider three different boundary conditions on the conductor:
\be
\left.
\Phi_{\omega}
\right|_{\bar{r}\to 1}
&=&0\ \ \ ({\rm D})\,,\\
\left.\frac{d}{d\bar{r}}
\left(
\frac{\Phi_{\omega}(\bar{r})}
       {\bar{r}}
\right)
\right|_{\bar{r}\to 1}
&=&
0
\ \ \ ({\rm N})\,,\\
\Phi'_{\omega}(1)&\to&
-i\bar{\omega} 
  \Phi_\omega (1)
\ \ \ ({\rm R})\,.
\ee
For the boundary condition (R),
Eq. \eqref{eq:S deformation} reduces to  
\be
\bar{\omega}^{2}
\int_{1}^{\infty}d\bar{r}
\left|
\Phi_{\omega}
\right|^{2}
+i\bar{\omega}
\left(
\left|\Phi_{\omega} (\infty)\right|^2
+
\left|\Phi_{\omega} (1)\right|^2
\right)
=
\int^{\infty}_{1}d\bar{r}
\left[
\left|
\Phi_{\omega}'
\right|^{2}
+
V_{\rm eff}(\bar{r})\left|
\Phi_{\omega}
\right|^{2}
\right]
\,.
\ee
Since the right-hand side is real, the imaginary part of the left-hand side has to vanish,
i.e.,  
\be
\bar{\omega}_R
\left(
2\bar{\omega}_I 
\int_{1}^{\infty}
d\bar{r}
\left|
\Phi_{\omega}
\right|^{2}
+
\left(
\left|\Phi_{\omega} (\infty)\right|^2
+
\left|\Phi_{\omega} (1)\right|^2
\right)
\right)
=0,
\ee
and we can show that $\bar{\omega}_{R}$ vanishes for unstable modes with $\bar{\omega}_{I}>0$.
For the boundary condition (D),
a similar analysis leads to 
\be
\label{integration_D}
\bar{\omega}_R
\left(
2{\bar\omega}_I 
\int_{1}^{\infty}
d\bar{r}
\left|
\Phi_{\omega}
\right|^{2}
+
\left|\Phi_{\omega} (\infty)\right|^2
\right)
=0,
\ee
and again $\bar{\omega}_{R}=0$ for $\bar{\omega}_{I}>0$.
Finally, for the boundary condition (N), 
Eq. \eqref{eq:S deformation} reduces to  
\be
\bar{\omega}^{2}
\int_{1}^{\infty}d\bar{r}
\left|
\Phi_{\omega}
\right|^{2}
+i \bar{\omega}
\left|\Phi_{\omega} (\infty)\right|^2
=
\left|
\Phi (1)
\right|^2
+
\int^{\infty}_{1}d\bar{r}
\left[
\left|
\Phi_{\omega}'
\right|^{2}
+
V_{\rm eff}(\bar{r})\left|
\Phi_{\omega}
\right|^{2}
\right]
\,.
\ee
Since the right-hand side is real,
a similar analysis leads to the same condition as \eqref{integration_D},
and hence shows that $\bar{\omega}_{R}=0$ vanishes for $\bar{\omega}_{I}>0$.
Thus, for all the boundary conditions considered,
we have shown that the eigenvalues of the unstable modes are always purely imaginary: $\bar{\omega}_R=0$.


\bibliography{lettersk}

 
\end{document}